\documentclass[traditabstract]{aa}
\usepackage{graphicx}
\usepackage{enumerate}
\usepackage{epsfig}
\usepackage{color}
\usepackage{txfonts}
\usepackage{natbib}
\usepackage{float}
\restylefloat{figure}

\bibpunct{(}{)}{;}{a}{}{,} 

\begin{document}

\title{A new look at the kinematics of the bulge from an N-body model}
\titlerunning{A new look at the kinematics of the bulge from an N-body model}

\author{A. G\'omez\inst{1},  P. Di Matteo\inst{1}, N. Stefanovitch, M. Haywood\inst{1}, F. Combes\inst{2}, D. Katz\inst{1}, C. Babusiaux\inst{1}}

\authorrunning{G\'omez et al.}

\institute{GEPI, Observatoire de Paris, CNRS, Universit\'e
  Paris Diderot, 5 place Jules Janssen, 92190 Meudon, France\\
\email{ana.gomez@obspm.fr}
\and
LERMA, Observatoire de Paris, CNRS, 61 Av. de l$'$Observatoire, 75014 Paris, France
}

\date{Accepted, Received}
\abstract{By using an N-body simulation of a bulge that was formed via a bar instability mechanism, we analyse the imprints of the initial (i.e. before bar formation) location of stars on the bulge kinematics, in particular on the heliocentric radial velocity distribution of bulge stars. Four different latitudes were considered: $b=-4^\circ$, $-6^\circ$, $-8^\circ$, and $-10^\circ$, along the bulge minor axis as well as outside it, at $l=\pm5^\circ$ and $l=\pm10^\circ$. The bulge X-shaped structure comprises stars that formed in the disk at different locations. Stars formed in the outer disk, beyond the end of the bar, which are part of the boxy peanut-bulge structure 
may show peaks in the velocity distributions at positive and negative heliocentric radial velocities with high absolute values that can be larger than 100 $\rm km$ $\rm s^{-1}$, depending on the observed direction. In some cases the structure of the velocity field is more complex and several peaks are observed. Stars formed in the inner disk, the most numerous, 
contribute predominantly to the X-shaped structure and present different kinematic characteristics. They display a rather symmetric velocity distribution and a smaller fraction of high-velocity stars. The stellar stream motion, which is induced by the bar changes with the star initial position, can reach more than $40$ $\rm km$ $\rm s^{-1}$ for stars that originated in the external disk, depending on the observed direction. Otherwise it is smaller than approximately $20$ $\rm km$ $\rm s^{-1}$. In all  cases, it decreases from $b=-4^\circ$ to $-10^\circ$. Our results may enable us to interpret the  cold high-velocity peak observed in the APOGEE commissioning data, as well as the excess of high-velocity stars in the near and far arms of the X-shaped structure at $l$=$0^\circ$ and $b$=$-6^\circ$.
When compared with real data, the kinematic picture becomes more complex due to the possible presence in the observed samples of classical bulge and/or thick disk stars. Overall, our results point to the existence of complex patterns and structures in the bulge velocity fields, which are generated by the bar. This suggests that caution should be used when interpreting the bulge kinematics: the presence of substructures, peaks and clumps in the velocity fields is not necessarily a sign of past accretion events.}


\maketitle

\section{Introduction}

It is now well-established that the Milky Way bulge hosts a bar and shows a characteristic boxy/peanut (B/P) shape  \citep[e.g. ][]{stanek94, dwek95, binney97, babugil05, lopezcor05, ratten07, cao13, weg13}.
From their analysis of star counts of red clump (RC) stars along the bulge minor axis,  \cite{mcwillzoc10} and \cite{nataf10}  independently detect the presence of a double density peak at $|b|>5^\circ,$ which was interpreted as evidence of the Galactic bulge being X-shaped. The distance between the two magnitude peaks is nearly constant with longitude and decreases toward the Galactic plane. These results were confirmed later by constructing 3D bulge density distributions using RC stars as distance indicators \citep{saito11, weg13}.  More recently, chemical abundances that were derived from a large number of RC bulge stars show that the double RC feature is only observed for the relatively metal-rich populations \citep{uttenthaler12, ness12, rojas14}. This conclusion has been questioned by \cite{nataf14}, who show a possible bias of metallicity in the determination of the distance of RC stars.

In addition to the shape of the Milky Way bulge, kinematic studies also support the existence of a bar. The Galactic bar produces kinematic distortions of the stellar velocity field since, in a pattern-rotating barred potential, the closed orbits are no longer circular but rather elongated along, or perpendicular to, the bar.
\cite{zhao94} and \cite{soto07}, analysing the space velocity of a sample of giant stars, obtain a significant vertex deviation that is consistent with a bar-like structure. Moreover, \cite{zhao94}, \cite{soto07}, and \cite{babusiaux10} show that the vertex deviation is mainly observed for the most metal-rich stars. 
Analysis of radial velocity measurements has given some insights into the kinematics  and the formation mechanisms of the Milky Way bulge. Thus the BRAVA M-giants survey \citep{rich07} shows that the bulge has cylindrical rotation \citep{how08,how09,kunder12}. This result was confirmed by more recent stellar bulge surveys using RC targets: the ARGOS survey \citep{ freeman13, ness13a, ness13b} and the GIBS survey \citep{zoccali14}, suggesting that a simple B/P bulge model should be able to reproduce the bulge kinematics \citep{shen10}. \\
On the other hand, stellar stream motions induced by the bar \citep{mao02} were detected by analysing the average radial velocities and/or proper motions of RC stars at the near and far sides of the bar. The results obtained were not always consistent with each other. Using radial velocity data, \cite{rangwala09} observe a modulus of the shift between the stellar streams ${\gtrsim30}$$\pm11$ $\rm km$ $\rm s^{-1}$ at $l=\pm5^\circ$ and $b=-3.5^\circ$. Analysing the radial velocities and metallicities of stars in the Galactic
plane at $l=\pm6^\circ$, \cite{babusiaux14} find streaming motion of stars and show that the highest velocity components are observed for metal-rich stars. Along the bulge minor axis, based on the analysis of radial velocities of stars in the double clump feature,  \cite{depropris11} at  $l=0^\circ$ and $b=-8^\circ$ and \cite{uttenthaler12} at $l=0^\circ$ and $b=-10^\circ$ find similar mean radial velocities and velocity dispersions  in the two RCs.
\cite{vasquez13}, comparing the two magnitude peaks of RC stars at $l=0^\circ$ and $b=-6^\circ$, detect a mean radial velocity difference in modulus of $\sim23$$\pm10$ $\rm km$ $\rm s^{-1}$. More recently, \cite{rojas14}, using radial velocity and metallicity data from the iDR1 Gaia-ESO survey \citep[see][]{gil12, randich13}, analyse the double RC kinematics and the stream motions in two fields: $l=0^\circ$ and $b=-6^\circ$ and $l=-1^\circ$ and $b=-10^\circ$ for relatively metal-rich stars. They obtain results compatible with those of  \cite{vasquez13}  and \cite{uttenthaler12} for the corresponding analysed regions. In Baade's window, using radial velocity data, \cite{babusiaux10} measure streaming motion of stars while \cite{rangwala09} find a radial velocity shift that is consistent with zero. Regarding proper motions data, \cite{babusiaux10}, using OGLE-II proper motions, do not find any statistically significant differences for stars in Baade's window while \cite{sumi03}, based on a larger sample, measure a shift between the two arms of the X-shaped structure. From OGLE-III proper motions, \cite{poleski13}, observe significant changes in the proper motion differences between the closer and the further arms of the X-shaped structure for $l> -0.1^\circ$ and $b=-5^\circ$, which are interpreted as the signature of the asymmetric streaming motion of stars.\\
Recently, \cite{nid12} report a cold high-velocity peak in the $V_{GSR}$ velocity distribution from the Apache Point Observatory Galactic Evolution Experiment (APOGEE) commissioning data. This result has been debated by \cite{li14} who, using N-body models, do not find a prominent peak with small velocity dispersion and argue that a spurious high-velocity peak may appear when the number of observed stars is limited. However, the simulation analysed by \cite{aumer15} reproduces the high-velocity peak and suggests that it is made up preferentially of young bar stars.

From chemical and kinematic studies there is evidence that the Milky Way bulge consists of several stellar populations that have different characteristics and origins \citep{babusiaux10,  gonzalez11, hill11, uttenthaler12, zoccali14, ness13b, rojas14}. There is consensus that  the relatively metal-rich stars support the B/P structure and are formed via internal evolution of the thin disk. Metal-poor stars, on the contrary, do not seem to be part of the peanut feature and their origin is controversial: an old spheroid that was formed via mergers or some dissipative collapse at early phases of the Galaxy formation (see the above cited studies), or a thick disk \citep{ness13b, dimatteo14, dimatteo15}. 

N-body dynamical models are able to qualitatively reproduce X-shaped bulges: a dynamically cold disk develops a bar from internal instability, which subsequently  buckles and heats the disk in the vertical direction giving rise to the typical B/P shape \citep[e.g. ][]{combes81, raha91, pfenniger91, athanassoula05, debattista06, martinez06,  li12, dimatteo14}. Several works have suggested that a pure disk instability model is able to explain reasonably well the observed characteristics of the Milky Way bulge, without the addition of any significant old spheroid or thick disk population \citep{shen10, martinez11, kunder12, martinez13, vasquez13, gardner14, zoccali14}. On the other hand, \cite{dimatteo14} show that a large part of the stellar disk, from the innermost regions to the outer Lindblad resonance (OLR), participates in the B/P-shaped bulge structure. As a result of the radial mixing, occurring at the time of the bar formation, stars that were initially distributed all over the disk are able to reach the bar region before its vertical buckling.  Moreover, the stars formed at large initial positions carry large angular momentum.  As a consequence the stellar bar is the result of a mixing of various stellar populations with different kinematic characteristics. By comparing the obtained results with the properties of the Milky Way's bulge  stellar populations that were observed by the ARGOS survey, \cite{dimatteo14} conclude that only the rich and moderately rich stellar populations, respectively called A and B in the ARGOS survey, are formed in the disk and are involved in the formation of the B/P structure.\\
As mentioned above, N-body simulations of a thin disk that has undergone a bar instability, are able to explain well enough the observed global structural and kinematic properties of the Milky Way bulge like the rotation curve, the velocity dispersion and the general trend of the velocity distribution, as obtained from radial velocity data. However, \cite{dimatteo15} show that the bulge of our Galaxy cannot be a pure thick stellar bar formed from a pre-existing thin stellar disk because in such a scenario, the detailed kinematic-chemistry relations, which were found by \cite{ness13b}, are not reproduced. \cite{dimatteo14} study the global dependence of the stellar kinematics on the star initial location on the Galactic plane and show that stars that originated in the inner disk exhibit an almost cylindrical rotation, while stars born in the outer disk do not. \cite{gardner14} and more recently \cite{quin15} explore the X-shape kinematics. In particular,  \cite{gardner14} show that the kinematic imprint of the X-shape is observed in the mean radial velocity difference between the closer and the further sides of the bar/bulge, but no coherent signature is found in Galactocentric azimuthal velocities, vertical velocities, or any of the velocity dispersions. 

Using an N-body simulation of a bulge that was formed via a bar instability mechanism already presented in \cite{dimatteo15}, the aim of this paper is to analyse the imprints of the stars' birth radii (distances in the Galactic plane with respect to the Galactic center at the beginning of the simulation) on the bulge kinematics, particularly on the heliocentric radial velocity distribution of bulge stars at different lines of sight. Because, at different observed directions, the line of sight should cross different velocity structures, we examine the trend of the variation of the mean heliocentric radial velocity field with distance. Instead of heliocentric distance, we use K-magnitude which, for RC stars, is a proxy of the heliocentric distance. The resulting velocity field is rather complex. It strongly depends on the stars' birth radii. Stars coming from the external regions of the disk, beyond the end of the bar up to the OLR, show an X-shaped structure and mainly contribute to the velocity distributions in the bulge at large distances from the Galactic centre. Stars formed in the inner parts of the disk, the most numerous, contribute predominantly to the X-shaped structure and are present at all distances in the bulge. Stars coming from the galactic external regions show, on average, higher radial velocities than stars that formed in the galactic inner regions. Our results show the existence of structures and/or high-velocity peaks in the velocity distributions, depending on the observed direction. They enable us to better understand the existing observations and to make predictions for future observational data.
 
The paper is organized as follows. In Section 2, our model is briefly described. Section 3 shows the imprints of the stars' birth radii in the X-shaped structure at different latitudes and longitudes. In Section 4, we perform the kinematic analysis. We first search the signatures that are due to the presence of non-circular motions on the global kinematics from Galactocentric radial and tangential velocities and their variation with the stars' birth radii. Then we investigate the consequences of the recovered structures on the 2D density distribution (heliocentric radial velocity versus K-magnitude, or distance) for the whole sample and for samples at different lines of sight. In Section 5, we argue that the kinematics of stars formed in the external disk may give a new interpretation of some results found in the literature, namely the stellar stream motions that are due to the bar and the presence of high-velocity stars in the bulge that are reported in the works of \cite{vasquez13} and \cite{nid12}. Finally, in Section 6, we draw our conclusions. 
 
\section{A Milky Way-like B/P-shaped model}\label{simall}

\begin{figure*}
\centering
\includegraphics[width=4.5cm,angle=0]{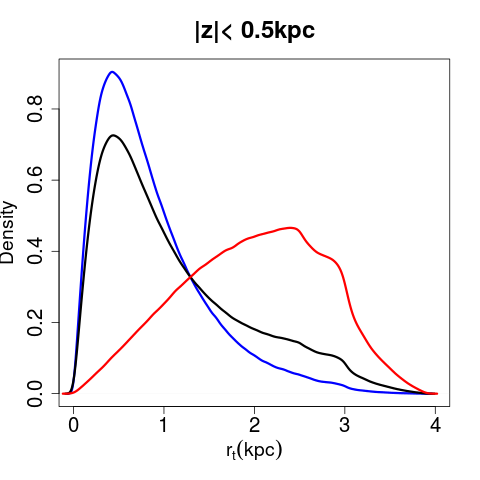}
\includegraphics[width=4.5cm,angle=0]{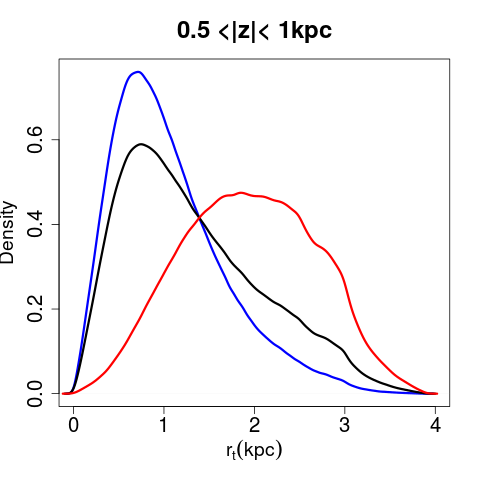}
\includegraphics[width=4.5cm,angle=0]{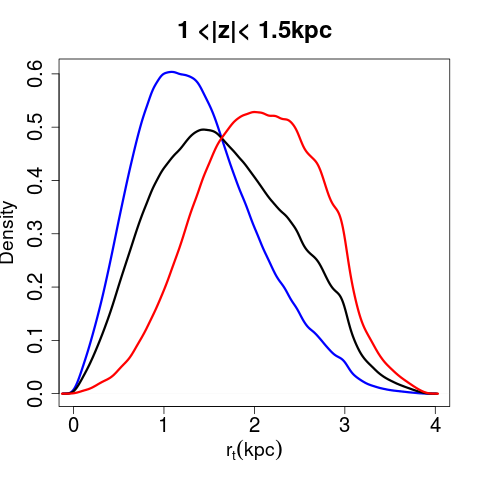}
\includegraphics[width=4.5cm,angle=0]{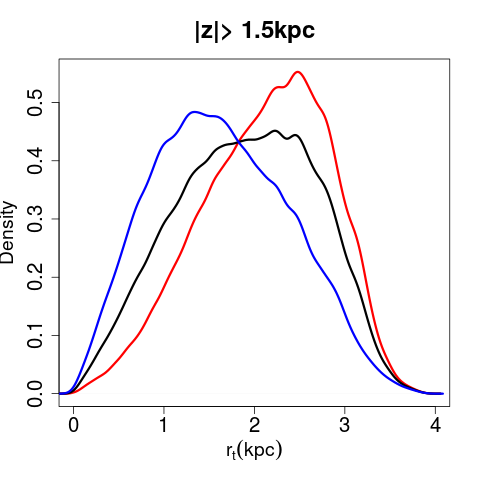}
\hspace{10pt}
\includegraphics[width=4.5cm,angle=0]{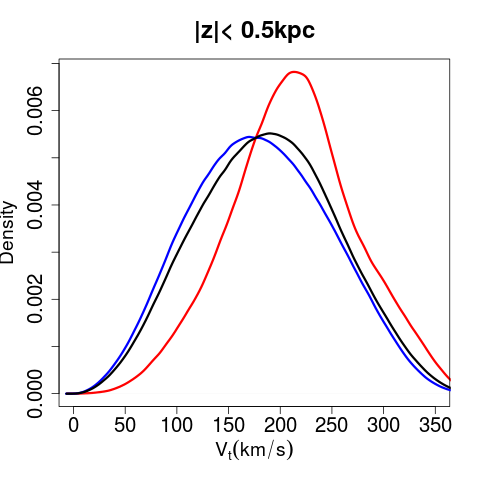}
\includegraphics[width=4.5cm,angle=0]{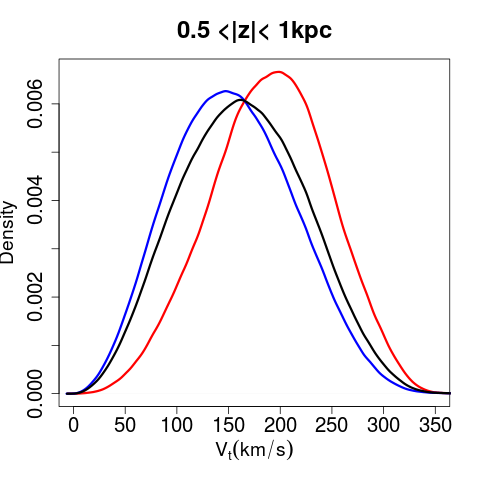}
\includegraphics[width=4.5cm,angle=0]{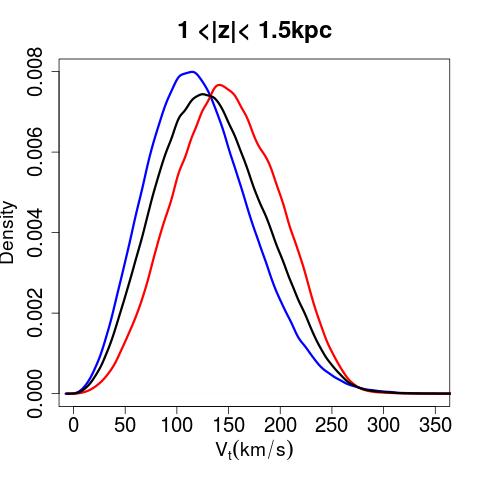}
\includegraphics[width=4.5cm,angle=0]{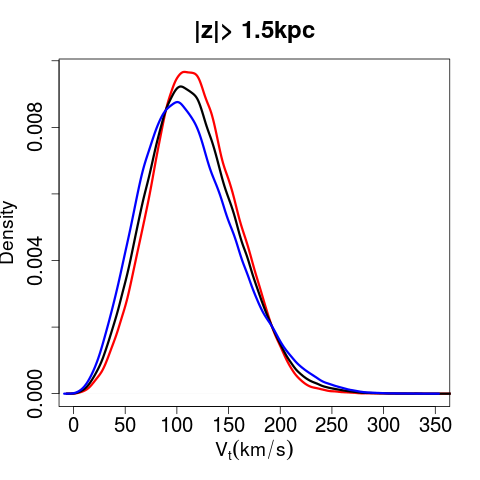}
\caption{First row: density distribution of $r_t$ (distance in the Galactic plane with respect to the Galactic centre at the final time of the simulation) of bulge stars at different slices in $|z|$ for the whole sample (in black) and for stars at birth radii $r_{ini}$ $<$ 2.5 kpc (in blue) and $r_{ini}{\ge 2.5}$ kpc (in red). Second row: density distribution of $V_t$ (total Galactocentric space velocity at the final time of the simulation) of bulge stars at different slices in $|z|$ for the whole sample (in black) and for stars at birth radii $r_{ini}$ $<$ 2.5 kpc (in blue) and $r_{ini}{\ge 2.5}$ kpc (in red).  Only disk stars with $|x|\le 2.5$~kpc, $|y|\le 3$~kpc, and $|z|\le 5$~kpc were selected.}
\label{Histo_rt_Vt}
\end{figure*}

In this paper, we analyse the high resolution simulation already described and analysed in \cite{dimatteo15}.  
It consists of an isolated stellar disk, with a B/D=0.1 classical bulge, and containing no gas. The choice of the adopted bulge-to-disk ratio (B/D) is consistent with the upper limit suggested for any classical bulge in the Milky Way  \citep{shen10, kunder12, dimatteo14}.\\
The dark halo and the bulge are modelled as Plummer spheres \citep{BT87}. The dark halo has a mass $M_H = 1.02 \times 10^{11} M_{\odot}$ and a characteristic radius $r_H=10$~kpc. The bulge has a mass $M_B = 9 \times 10^9M_{\odot}$ and a  characteristic radius $r_{B}=1.3$~kpc. The stellar disk follows a Miyamoto-Nagai density profile \citep{BT87}, with mass $M_* = 9 \times 10^{10} M_{\odot}$ and vertical and radial scale lengths given by $h_* =0.5$~kpc and  $a_* =4$~kpc, respectively. The initial disk size is 13~kpc, and the Toomre parameter is set equal to Q=1.8. The galaxy is represented by $N_{tot}$= 30 720 000 particles redistributed among dark matter ($N_H$= 10 240 000) and stars ($N_{stars}$ = 20 480 000). To initialize particle velocities, we adopted the method described in \citet{hern93}. A Tree-SPH code \citep{sem02} was used to run the simulations.  A Plummer potential is used to soften gravity at scales smaller than $\epsilon$ = 50 pc. The equations of motion are integrated over 4 Gyr, using a leapfrog algorithm with a fixed time step of $\Delta t = 2.5 \times 10^4$ yr. In this work, we analyse a snapshot of this simulation in particular, corresponding to a time of about 3 Gyr from the beginning of the simulation. At this epoch, the bar has already acquired its B/P-shaped morphology.\\
The simulation was rescaled to match the Milky Way bar size and bulge velocities \citep[see][]{ dimatteo15}, the bar semi-major axis length ($r_{bar}$) being ${\sim3.5}$ kpc, the corotation is at 4.5~kpc, the OLR at 7.5~kpc, and the Sun position at 8~kpc from the Galactic centre. For consistency, we adopted a bar orientation relative to the Sun-Galactic center of 20$^\circ$, as in \cite{dimatteo15}.
To avoid contamination from background and foreground stars, we selected stars inside the bulge region, defined by $|x|\le 2.5$~kpc and $|y|\le 3$~kpc, where x and y are measured on the Galactic plane from the Galactic center, y in the direction Sun-Galactic center, and x perpendicular to it. The axis perpendicular to the Galactic plane is z.\\ 
\cite{dimatteo14} discuss in detail how the presence of the stellar bar affects the spatial redistribution of stars in the disk. According to their birth radii ($r_{ini}$), defined as the distance in the Galactic plane with respect to the Galactic centre at the beginning of the simulation, they showed that stars that formed inside the inner Lindblad resonance (about 1.5 kpc), mainly remain confined in the inner bar region, and that those formed in the outer disk (beyond the corotation) migrate both outward and inward, reaching both the edges (up to the OLR) and the centre of the disk (see their Fig. 3). Furthermore, the spatial redistribution of stars in the disk after the bar formation, which is a consequence of the angular momentum redistribution, show that, on average, the final angular momentum is higher for stars with larger birth radii (see their Fig. 5). Using our simulation, Fig.~\ref{Histo_rt_Vt} (first row) displays for disk stars the density distribution of $r_t$ for slices of $|z|$, $r_{t}$  being the distance in the Galactic plane with respect to the Galactic centre at the final time of the simulation. In each plot, two different birth radii slices were considered: $r_{ini}$ $< $ 2.5kpc (inner disk formation) and $r_{ini}$ $>$ 2.5kpc (external disk formation). As we can see, stars that formed everywhere in the disk, from the center to the ORL, are present in the bulge. The number of stars formed in the external disk regions increases with the height from the plane. Furthermore, in the outer bar region, this number is more significant compared to the number of stars formed in the innermost disk. Regarding the kinematics, at the beginning of the simulation, the stars move in almost circular orbits. After the bar formation, the star orbits are significantly affected.
Figure~\ref{Histo_rt_Vt} (second row) shows the density distribution of the total Galactocentric space velocity at the final time of the simulation ($V_t$) for the same slices of $|z|$. In each plot the same slices in $r_{ini}$ were considered. We note that stars that come from the galactic external regions, show, on average, higher total Galactocentric space velocities than stars formed in the galactic inner regions. In the following sections, we analyse the imprints of the star birth radii on the spatial distribution and on the kinematics of bulge stars at different lines of sight.
%
Four different latitudes were considered: $b=-4^\circ$, $-6^\circ$, $-8^\circ$, and $-10^\circ$ along the bulge minor axis, as well as outside it, at $l=\pm5^\circ$ and $l=\pm10^\circ$. \\
%
\begin{figure*}
\centering
\includegraphics[width=16.8cm,angle=0]{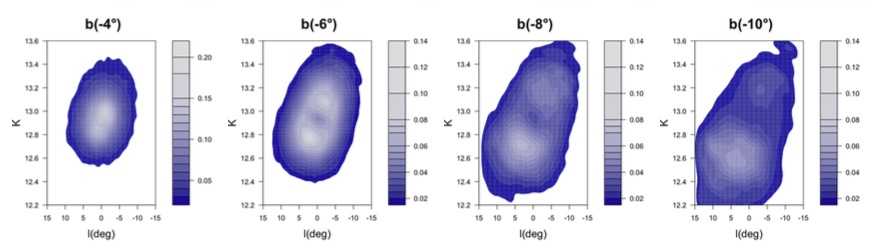}
\caption{K-magnitude-Galactic longitude $l$ density distribution contour plots of the N-body model for different latitudes $b=-4^\circ$, $-6^\circ$, $-8^\circ$, and $-10^\circ$, $\Delta{b} =1^\circ$ }
\label{Kldistri}
\end{figure*}
Even though our numerical model was not designed to match the actual Galactic bulge, it reproduces rather well the main features of its structure and kinematics. In \cite{dimatteo15} (Fig. 1) the rotation curve and the radial velocity dispersions of the N-body model are compared to BRAVA and ARGOS data. As shown in this work, the model reproduces the overall kinematic observed trends well: the approximately cylindrical rotation and the decrease and flattening of the velocity dispersions with $b$. \\
Figure~\ref{Kldistri} shows the density maps of the apparent K-magnitude of RC stars versus Galactic longitude for the four considered latitude directions, the size of the fields being $ \Delta{l}=\Delta{b} =1^\circ$. K-magnitudes were calculated by adopting an absolute magnitude for the clump stars of $M_K$=-1.61 \citep{alves00}, which gives the minimum of the split clump at K =12.9 mag. The double peak structure is clearly visible and the separation between the two peaks increases from $b=-4^\circ$ to $-10^\circ$ as shown in the observed data \citep{mcwillzoc10, nataf10, weg13}. 
We note that, in our simulation, the split is also observed at $l=0^\circ$, $b=-4^\circ$ in agreement with \cite{weg13} results. However, the simulation does not reproduce the observed separation of the prominent peaks well enough. Our values are 25-30\% smaller than the measured separation in the direction near $l=0^\circ$, $b=-6^\circ$ \citep{mcwillzoc10, nataf15}. On the other hand, the observational identification of the double RC feature is affected by different uncertainties, such as sample contamination, extinction, errors in the colour-magnitude diagram, and selection effects. Moreover, as we can see in the next section, the contribution to the peaks depends on the stellar birth radius.

\section{Imprints of the stars' birth radii on the X-shaped structure}\label{impX}

Figure~\ref{Histo_rt_Vt} shows that the contribution to the boxy bulge of stars with different birth radii varies with $|z|$. Because, in different regions, the line of sight passes through different structures inside the peanut structure, the contribution of stars of different origins should vary at a given latitude with longitude.
\begin{figure*}
\centering
\includegraphics[trim = 0cm 0cm 0cm 0cm, width=0.7\textwidth,angle=0]{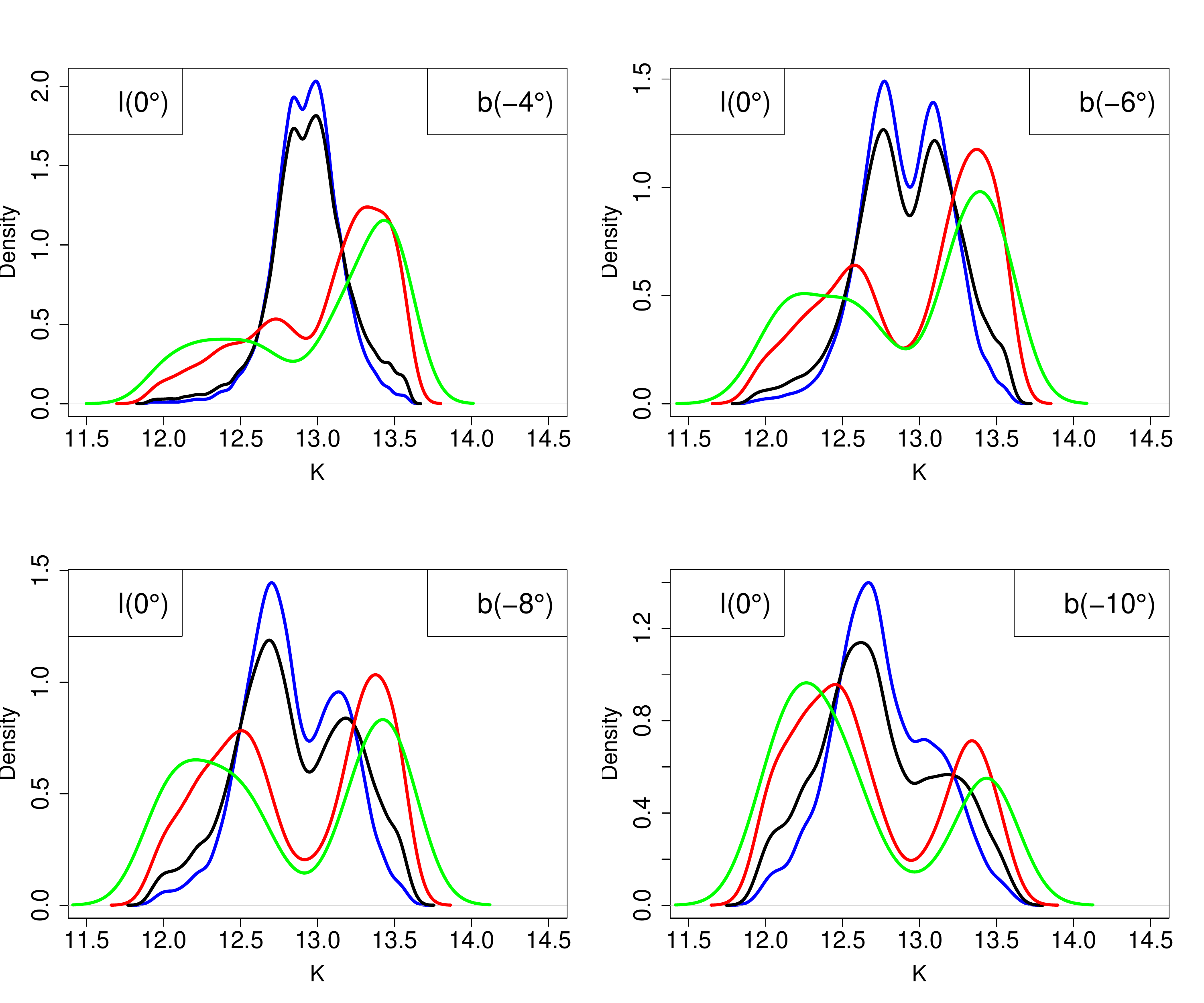}
\caption{K-magnitude density distribution of RC stars along the bulge minor axis. Only disk stars with $|x|\le 2.5$~kpc and $|y|\le 3$~kpc were selected. Four latitudes centred at $b$=$-4^\circ$, $-6^\circ$, $-8^\circ$, and $-10^\circ$ from top left to bottom right are shown. The size of the fields is $ \Delta{l}=\Delta{b} =1^\circ$ for  $b$ ${\geq}-6^\circ$ and $ \Delta{l}=\Delta{b} =1^\circ.5$ for $b$ $ < -6^\circ$. For each plot, all stars of the modelled galaxy are shown  (black lines), as well as  
 stars selected on their birth radii according to: $r_{ini}$ $<$ 2.5 kpc (blue lines), 2.5  kpc ${\le}r_{ini}$ $<$ 4.5 kpc (red lines), and  $r_{ini}{\ge4.5}$ kpc (green lines).  }
\label{Kl0bm}
\end{figure*}
The distributions of the apparent K-magnitude of RC stars obtained from the model for different birth radii $r_{ini}$ 
are displayed in Fig.~\ref{Kl0bm} for stars along the bulge minor axis and, in Fig.~\ref{Kl5bm}, and Fig.~\ref{Kl10bm} for the other lines of sight. The fraction of stars in the different selected regions is shown in Fig.~\ref{Pourcent_rini} for different $r_{ini}$.
Only disk stars were plotted because classical bulge stars remain concentrated towards the Galactic centre and do not participate in the peanut feature. Following \cite{dimatteo14}, three $r_{ini}$ values have been considered: 
$r_{ini}$ $<$ 2.5 kpc (in blue), 2.5  kpc ${\le}$ $r_{ini}$ $<$ 4.5 kpc (in red), and $r_{ini}$ ${\ge4.5}$ kpc (in green).  
In terms of the semi-major axis length of the bar, $r_{bar}$, these values correspond approximately to $r_{ini}$ $<$ 0.7$r_{bar}$, 
0.7$r_{bar}$ ${\le}$ $r_{ini}$ $<$ 1.3$r_{bar}$, and $r_{ini}$ ${\ge}$ 1.3$r_{bar}$, respectively.
The size of the fields is $ \Delta{l}=\Delta{b} =1^\circ$ for $b$ ${\geq}-6^\circ$. To get better number statistics for samples with $r_{ini}$ ${\ge4.5}$ kpc at latitudes $b$ $<-6^\circ$, $ \Delta{l}=\Delta{b} =1^\circ.5$ were adopted. 

\begin{figure*}
\centering
 \includegraphics[trim = 0cm 0cm 0cm 0cm, width=0.7\textwidth,angle=0]{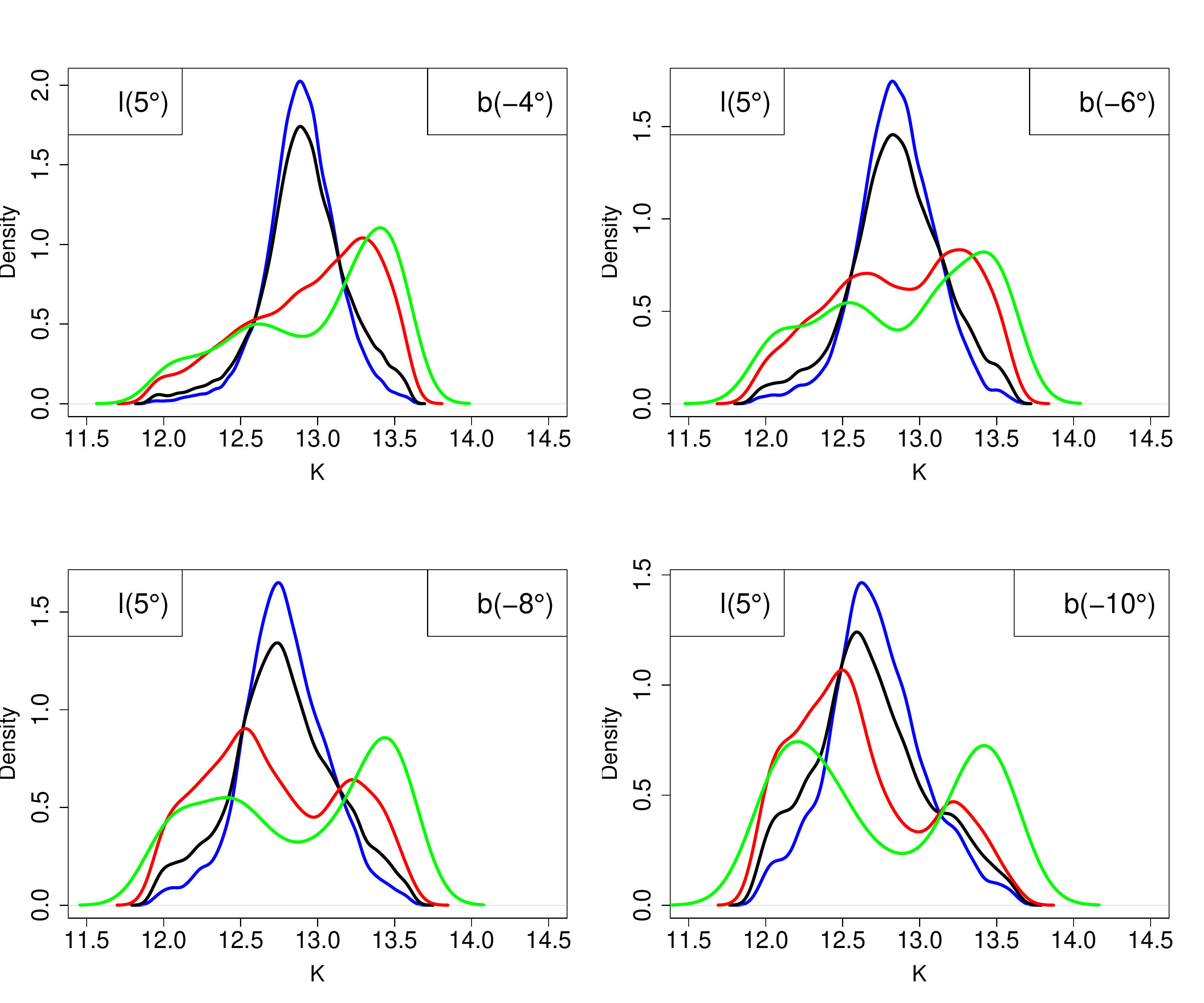}
 \includegraphics[trim = 0cm 0cm 0cm 0cm, width=0.7\textwidth,angle=0]{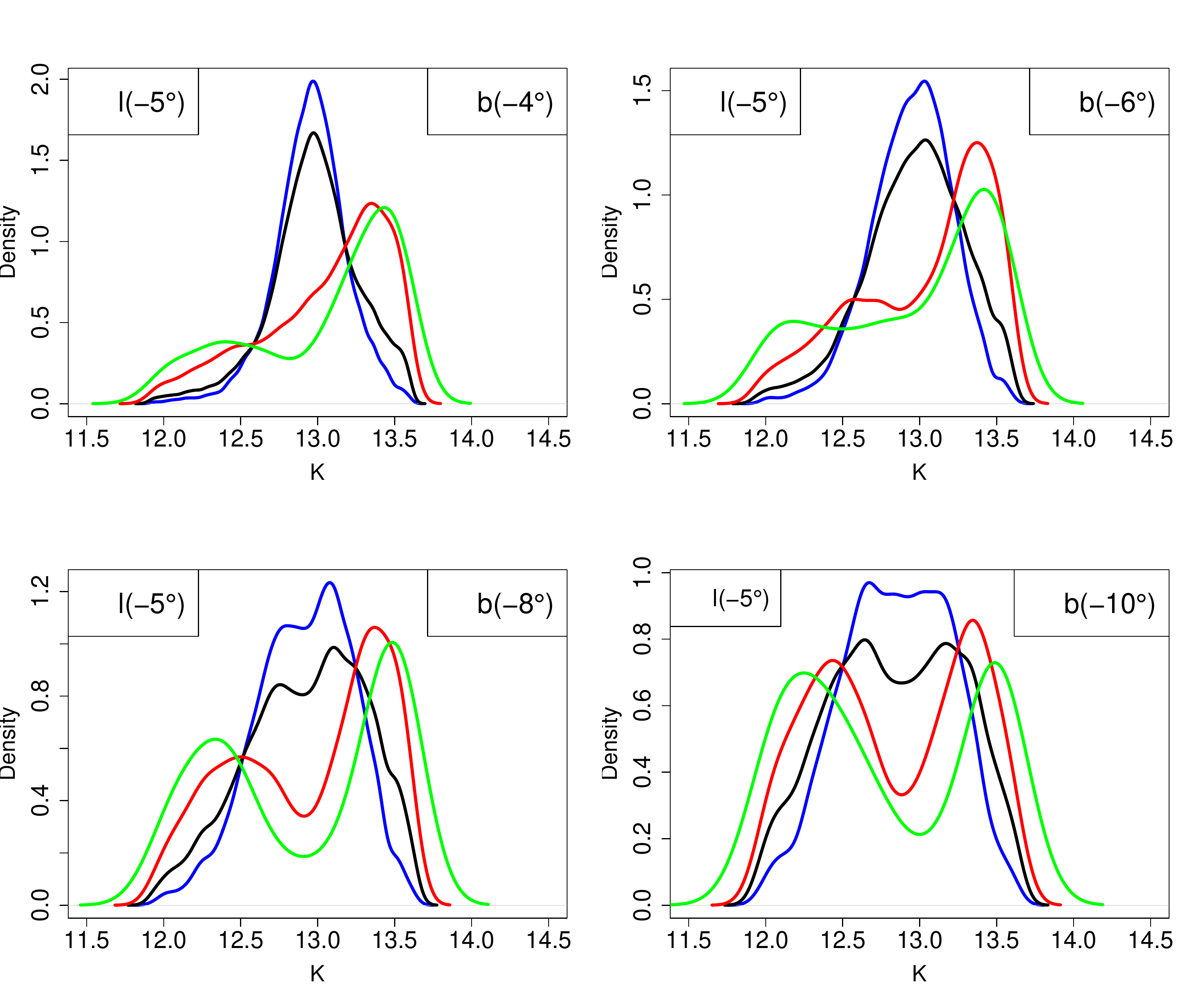}
 \caption{K-magnitude density distribution of RC stars outside the bulge minor axis centred at $l$=$5^\circ$ and $l$=$-5^\circ$. Only disk stars with $|x|\le 2.5$~kpc and $|y|\le 3$~kpc were selected. For each longitude, four latitudes centred at $b$=$-4^\circ$, $-6^\circ$, $-8^\circ$, and $-10^\circ$  from top left to bottom right are shown. The size of the fields is $ \Delta{l}=\Delta{b} =1^\circ$ for  $b$ ${\geq}-6^\circ$ and $ \Delta{l}=\Delta{b} =1^\circ.5$ for $b$ $ < -6^\circ$. For each plot, all stars of the modelled galaxy are shown (black lines), as well as  
 stars selected on their birth radii according to: $r_{ini}$ $<$ 2.5 kpc (blue lines), 2.5  kpc ${\le}r_{ini}$ $<$ 4.5 kpc (red lines), and  $r_{ini}{\ge4.5}$ kpc (green lines).}
\label{Kl5bm}
\end{figure*}
\begin{figure*}[!h]
 \centering
 \includegraphics[trim = 0cm 0cm 0cm 0cm, width=0.7\textwidth,angle=0]{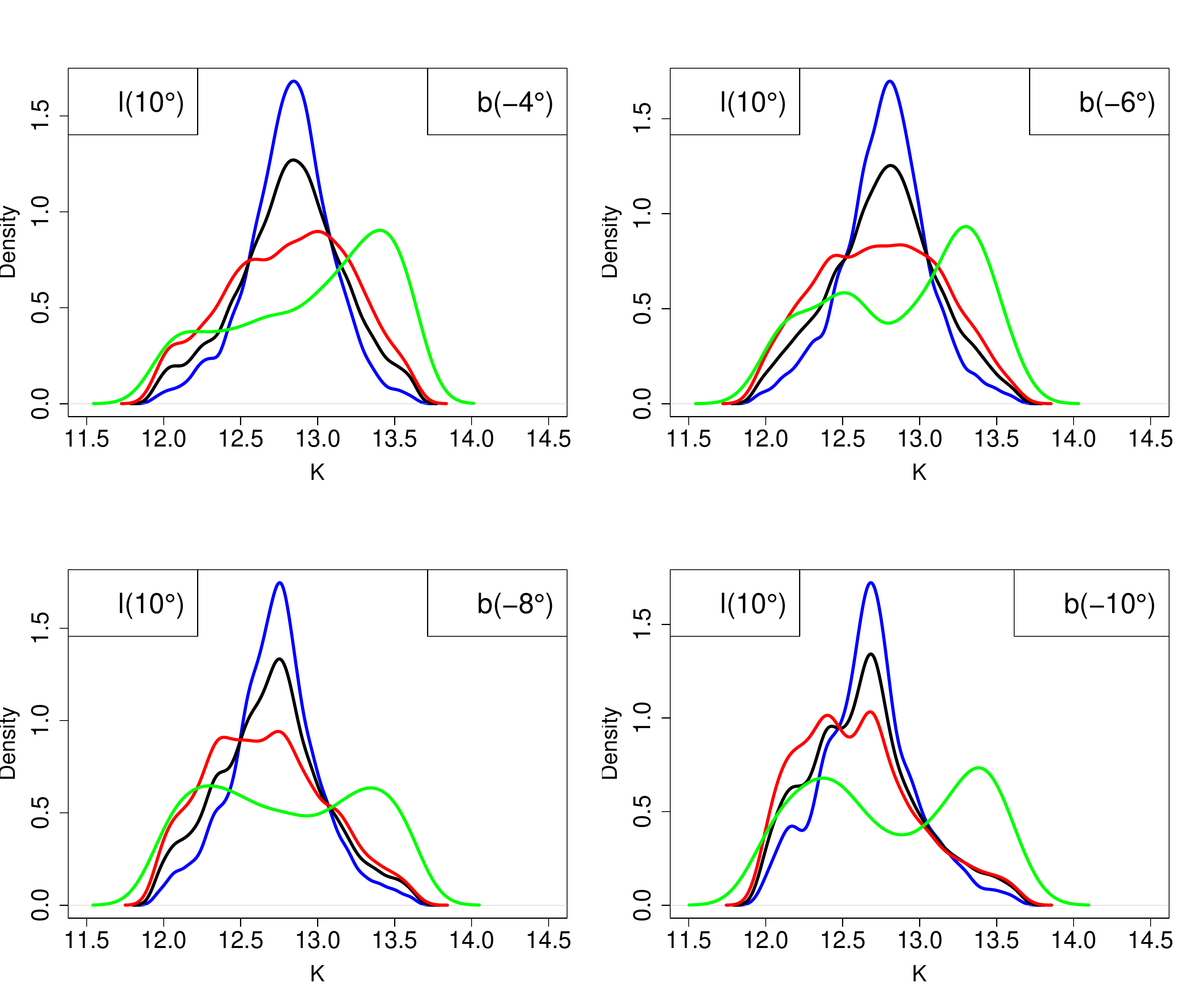}
 \includegraphics[trim = 0cm 0cm 0cm 0cm, width=0.7\textwidth,angle=0]{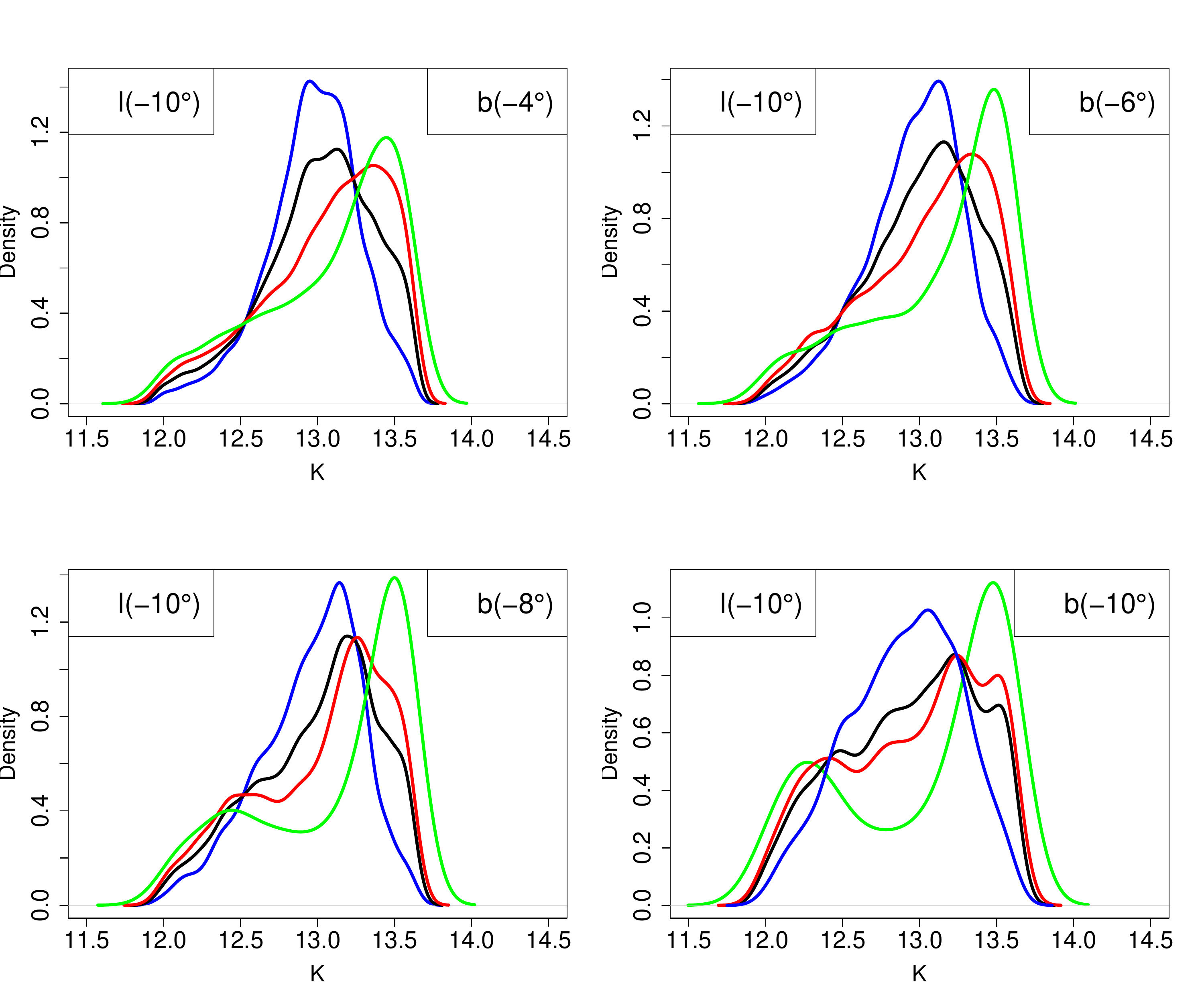} 
 \caption{K-magnitude density distribution of RC stars outside the bulge minor axis centred at $l$=$10^\circ$ and $l$=$-10^\circ$. Only disk stars with $|x|\le 2.5$~kpc and $|y|\le 3$~kpc were selected. For each longitude, four latitudes centred at $b$=$-4^\circ$, $-6^\circ$, $-8^\circ$, and $-10^\circ$  from top left to bottom right are shown. The size of the fields is $ \Delta{l}=\Delta{b} =1^\circ$ for  $b$ ${\geq}-6^\circ$ and $ \Delta{l}=\Delta{b} =1^\circ.5$ for $b$ $ < -6^\circ$. For each plot, all stars of the modelled galaxy are shown  (black lines), as well as  
 stars selected on their birth radii according to: $r_{ini}$ $<$ 2.5 kpc (blue lines), 2.5  kpc ${\le}r_{ini}$ $<$ 4.5 kpc (red lines), and  $r_{ini}{\ge4.5}$ kpc (green lines).}
\label{Kl10bm}
\end{figure*}
\begin{figure*}[!h]
\centering
\includegraphics[trim = 0cm 0cm 0cm 0cm, width=0.3\textwidth,angle=0]{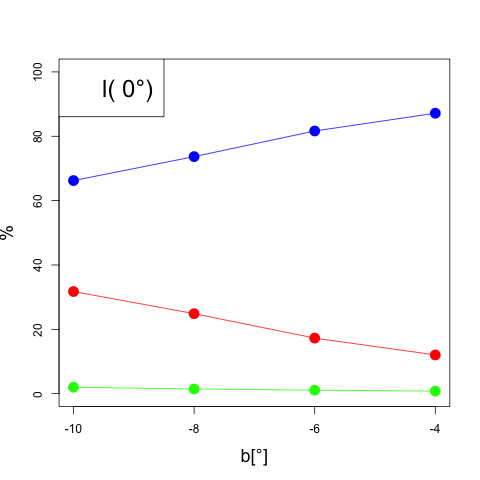}
\includegraphics[trim = 0cm 0cm 0cm 0cm, width=0.3\textwidth,angle=0]{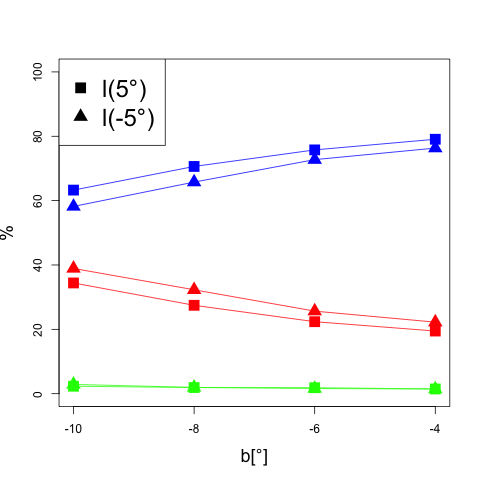}
\includegraphics[trim = 0cm 0cm 0cm 0cm, width=0.3\textwidth,angle=0]{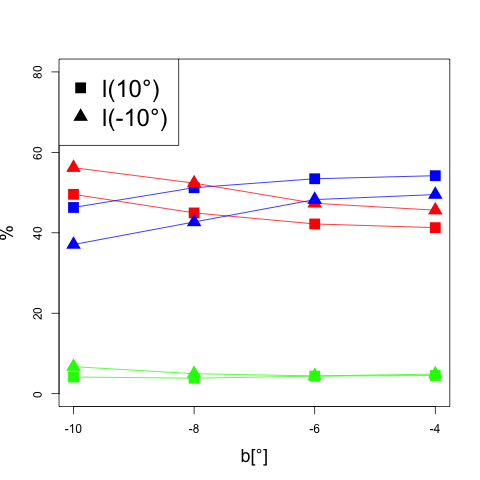}
\caption{Fraction of stars (in percentage) in the selected directions according to their birth radii: $r_{ini}$ $<$ 2.5 kpc (blue symbols), 2.5  kpc ${\le}r_{ini}$ $<$ 4.5 kpc (red symbols), and $r_{ini}{\ge4.5}$ kpc (green symbols) }
\label{Pourcent_rini}
\end{figure*}
Along the bulge minor axis, Fig.~\ref{Kl0bm} displays a significant bimodal K-magnitude distribution for stars at $b$ ${\le-6}^\circ$.  
All the stars contribute to the bimodality. At a given latitude, the separation between the nearest and the most distant clumps increases with the birth radius. Stars formed in the inner regions ($r_{ini}$ ${< 2.5}$ kpc) mainly contribute to the closer peaks. For stars born at larger distances ($r_{ini}$ ${\ge2.5}$ kpc), the peaks are further separated and contribute to the more external parts of the peanut. Furthermore, the separation between the peaks increases from $b=-4^\circ$ to $-10^\circ$. 
Figure~\ref{Pourcent_rini} (left) shows that the main contribution to the X-shaped arms comes from stars formed in the inner galactic disk ($r_{ini}$ $<$ 2.5 kpc). It varies from about 82$\%$ at $b=-6^\circ$ to 66$\%$ at $b=-10^\circ$. The fraction of stars that originated at $r_{ini}$ between 2.5 and 4.5 kpc increases from $b=-4^\circ$ (12$\%$) to $-10^\circ$ (32$\%$), while it is almost constant for stars that originated at $r_{ini}$ $\ge 4.5$ kpc (less than about 2$\%$).\\
 Outside the bulge minor axis, depending on the observed direction, the line of sight can cross the two X-shaped arms or only one. In Figs.~\ref{Kl5bm} and  \ref{Kl10bm}, stars with $r_{ini}$ $<$ 2.5 kpc show one peak centred at K-magnitude $< 13$ for positive $l$ while, for negative $l,$ the distributions are different. At $l=-5^\circ$, a slightly double feature is present at $b$ $<$ $-6^\circ$, while only one peak is observed at $b$ $\ge$ $-6^\circ$. At $l=-10^\circ$ the K-magnitude distributions do not display bimodality for stars with $r_{ini}$ $<$ 2.5 kpc.
However, for stars with $r_{ini}$ ${\ge2.5}$ kpc, in almost all of the investigated directions, the distributions exhibit a bimodality  that is especially noticeable at $b=-8^\circ$ and $b=-10^\circ$. Figure~ \ref{Pourcent_rini} (middle and right) shows that the contribution at $l=\pm5^\circ$ of stars formed between 2.5 and 4.5 kpc varies from about 30${\%}$ at $b=-8^\circ$ to 40${\%.}$ at $b=-10^\circ$. At $l=\pm10^\circ$, this fraction changes little with latitude, from about 42${\%}$ at $b=-6^\circ$ to 50${\%}$ at $b=-10^\circ$ at $l=10^\circ$, these values being slightly greater at $l=-10^\circ$. On the other hand, stars with $r_{ini}$ ${\ge4.5}$ kpc are a small fraction of all samples, less than 7${\%}$.\\ 
Even though the different contributions depend on the bar orientation with respect to the Sun-Galactic centre direction, our results clearly indicate that the prominent contribution to the X-shaped feature comes from stars formed in the inner regions. However,  stars that were formed in more external regions also show bimodality, the separation of the peaks increasing with the birth radius.  As a consequence the 3D shape of the X-shape should actually look like a peanut as has been already mentioned by \cite{shen15}.

\section{Imprints of the stars' birth radii on the global kinematics}
Since stars with different origins carry different angular momentum, the imprints of the stars' birth radii should be found on the observed kinematics of bulge stars (\cite{dimatteo14}, Fig.5). As shown in Fig.\ref{Histo_rt_Vt}, the density distribution of the total Galactocentric space velocity varies with the stars' birth radii.

\begin{figure*}
\centering
\includegraphics[width=16cm,angle=0]{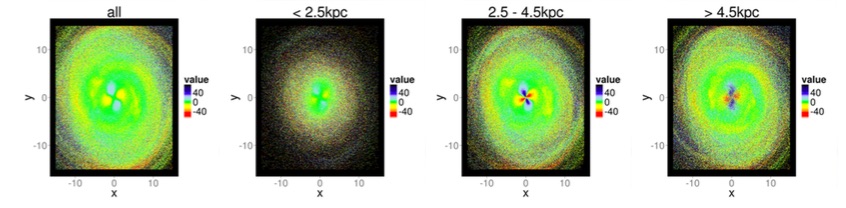}
\caption{Face-on maps of the $V_{\pi}$ velocity of stars born at different radii. Only disk stars with $|x|\le 15$~kpc, $|y|\le 15$~kpc, and $|z|\le5$~kpc are shown. From left to right: all stars of the modelled galaxy, stars at birth radii $r_{ini}$ $<$ 2.5 kpc, 2.5 kpc ${\le}r_{ini}$ $<$ 4.5 kpc, and $r_{ini}{\ge4.5}$ kpc. }
\label{vpi}
\end{figure*}

\begin{figure*}
\centering
\includegraphics[width=16cm,angle=0]{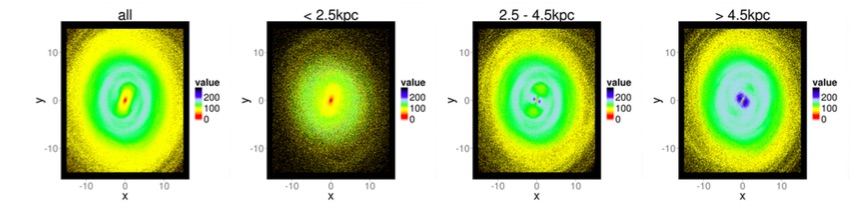}
\caption{Face-on maps of the $V_{\theta}$ velocity of stars born at different radii. Only disk stars with $
|x|\le15$~kpc, $|y|\le 15$~kpc, and $|z|\le5$~kpc are shown. From left to right: all stars of the 
modelled galaxy, stars at birth radii $r_{ini}$ $<$ 2.5 kpc, 2.5 kpc ${\le}r_{ini}$ $<$ 4.5 kpc, and $r_{ini
}{\ge4.5}$ kpc. }
\label{vtheta}
\end{figure*}

\begin{figure*}[h]
\centering
\includegraphics[width=16cm,angle=0]{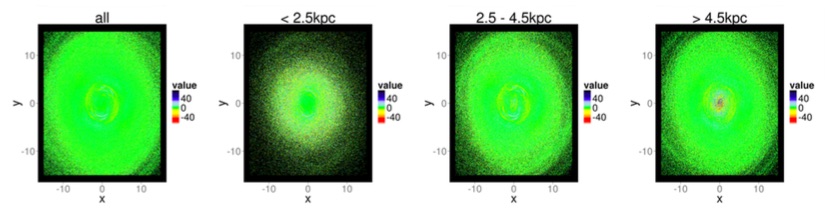}
\caption{Face-on maps of the $V_{z}$ velocity of stars born at different radii. Only disk stars with $|x|\le 15$~kpc, $|y|\le15$~kpc, and $|z|\le5$~kpc are shown. From left to right: all stars of the modelled galaxy, stars at birth radii $r_{ini}$ $<$ 2.5 kpc, 2.5 kpc ${\le}r_{ini}$ $<$ 4.5 kpc, and $r_{ini}{\ge4.5}$ kpc. }
\label{vz}
\end{figure*}

In this section, after analysing the Galactocentric velocity maps, we investigate the expected signatures of the velocity structures on the radial velocity component measured from the Sun  ($V_H$). Because  the line of sight crosses different velocity structures for different observed
directions, we examine the variation of $V_H$ with K-magnitude, which for RC stars is a proxy of the heliocentric distance. Moreover, as shown
in the previous section, because the morphology of the X-shaped structure depends on the star birth radius, we can also expect that  the variation of 
$V_H$ with K-magnitude depends on it. In the following, we investigate this point in detail.\\
Figures~\ref{vpi}, \ref{vtheta}, and \ref{vz} display, respectively, the face-on maps of the radial ($V_{\pi}$), tangential ($V_{\theta}$), and vertical ($V_{z}$) velocities in cylindrical Galactocentric coordinates ($R$, $\theta$, $z$), for all disk stars and for stars selected according to their birth radii. Only disk stars within $|x|\le15$~kpc, $|y|\le15$~kpc, and 
$|z|\le5$~kpc were included. The selection in $|z|$ was introduced to avoid stars at large distances. The presence of non-circular motions in the bulge region is clearly observed in the $V_{\pi}$  and $V_{\theta}$ components. Figure~\ref{vpi} shows that the presence of the bar induces peculiar structures in the Galactocentric radial velocity $V_{\pi}$ maps: four regions exist in the inner disk, two with positive $V_{\pi}$, indicating outward motions, and two with negative $V_{\pi}$, indicating inward motions.  The amplitude of these motions (in modulus) increases with $r_{ini}$. The maps of $V_{\theta}$ present contours parallel to the bar. For stars with $r_{ini}$ $>$ 2.5~kpc, Fig.~\ref{vtheta} shows a clear velocity tangential structure at the edge of the bar major axis, the amplitude of the corresponding motion increases with birth radius. Contrary to $V_{\pi}$ and $V_{\theta}$ velocity distributions, a clear structure in the global distribution of $V_{z}$ velocities is not observed, as expected. 

%

%

\subsection{Heliocentric radial velocity -- K-magnitude density distributions in the bulge}
\begin{figure*}
\centering
\includegraphics[width=4.0cm,angle=0]{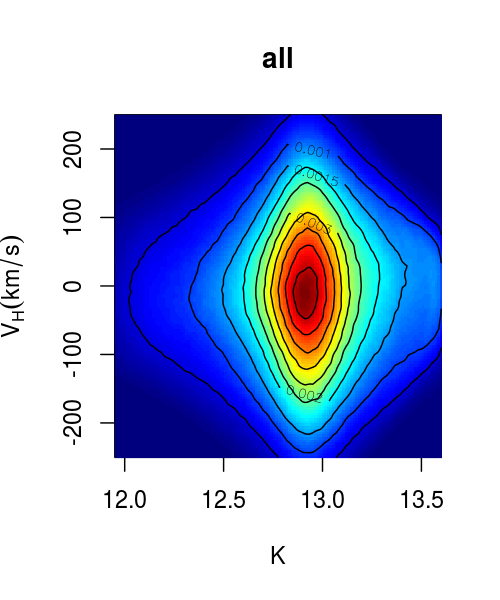}
\includegraphics[width=4.0cm,angle=0]{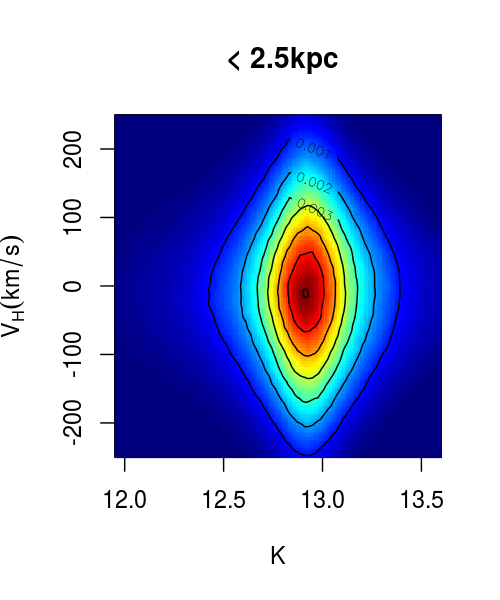}
\includegraphics[width=4.0cm,angle=0]{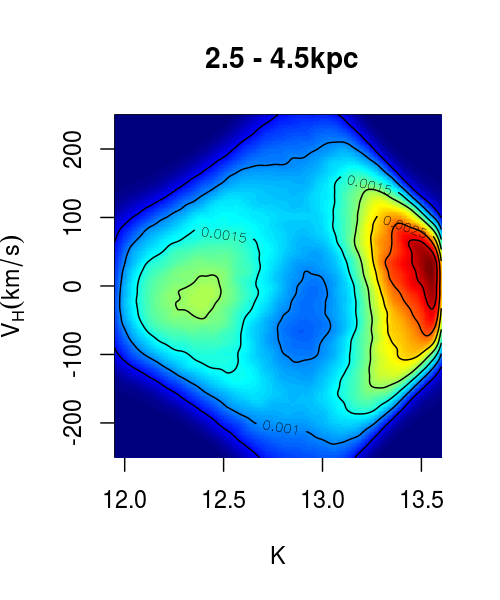}
\includegraphics[width=4.0cm,angle=0]{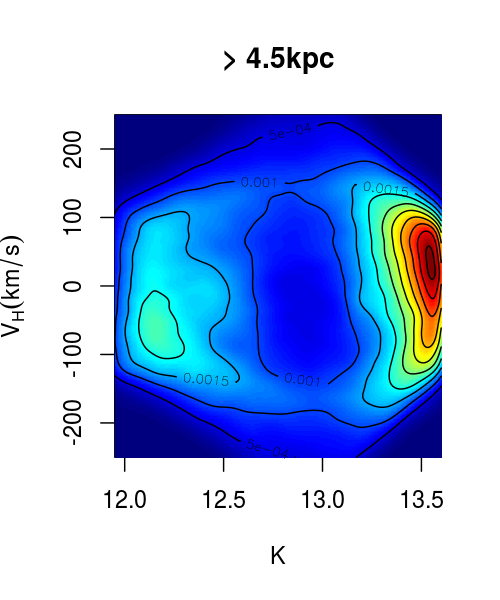}
\hspace{10pt}
\includegraphics[width=4.0cm,angle=0]{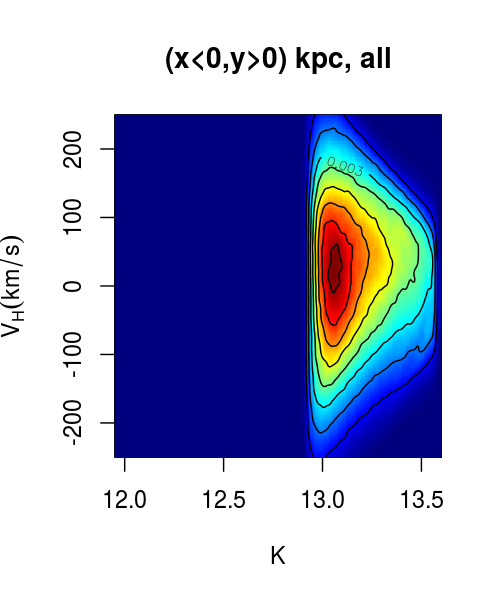}
\includegraphics[width=4.0cm,angle=0]{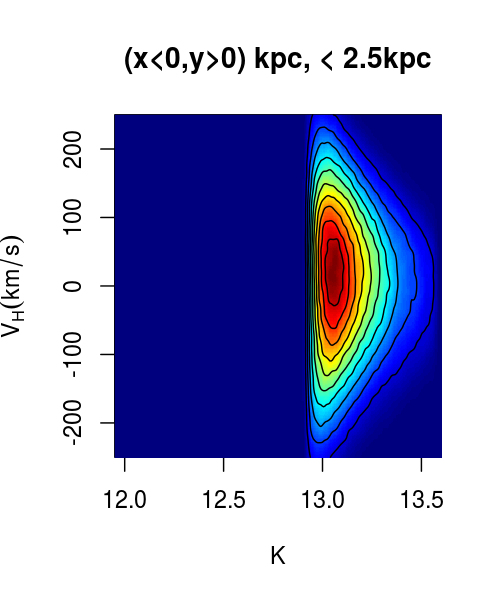}
\includegraphics[width=4.0cm,angle=0]{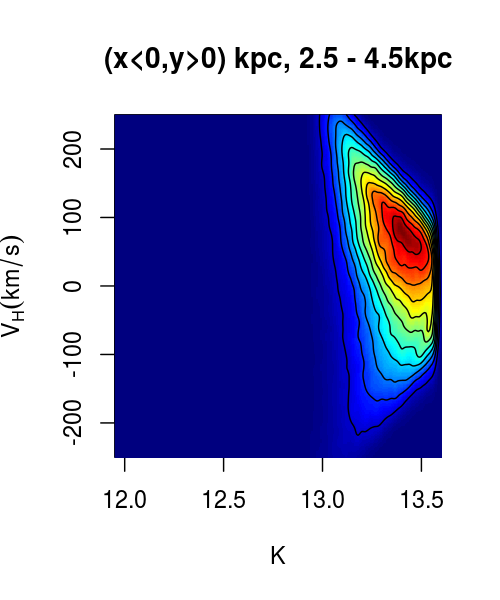}
\includegraphics[width=4.0cm,angle=0]{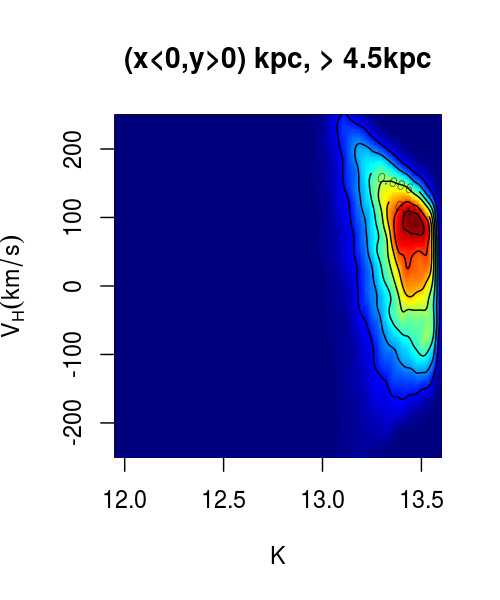}
\hspace{10pt}
\includegraphics[width=4.0cm,angle=0]{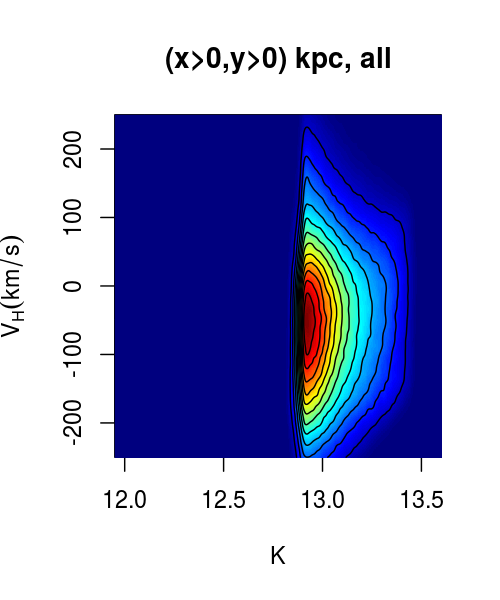}
\includegraphics[width=4.0cm,angle=0]{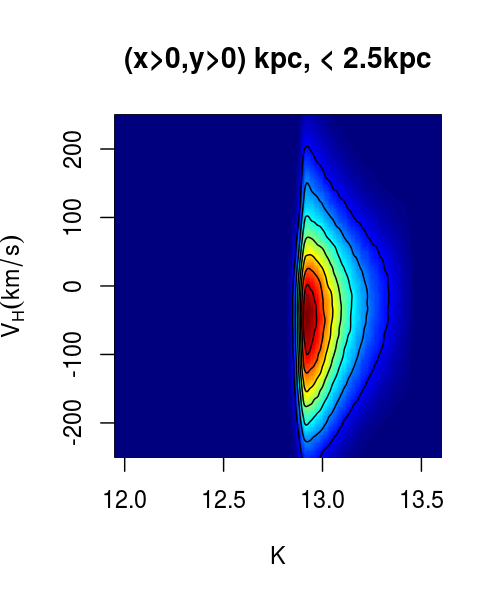}
\includegraphics[width=4.0cm,angle=0]{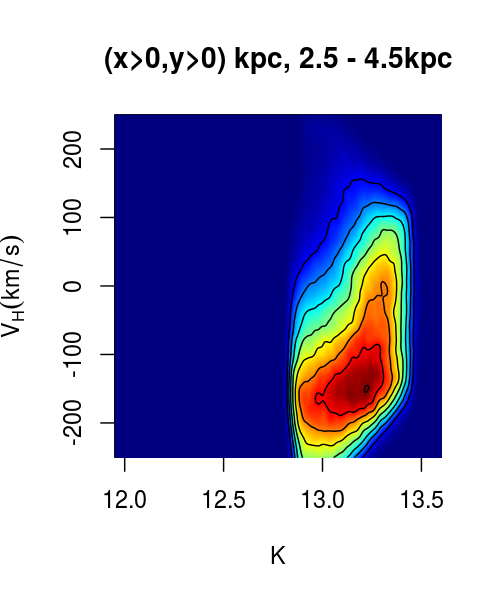}
\includegraphics[width=4.0cm,angle=0]{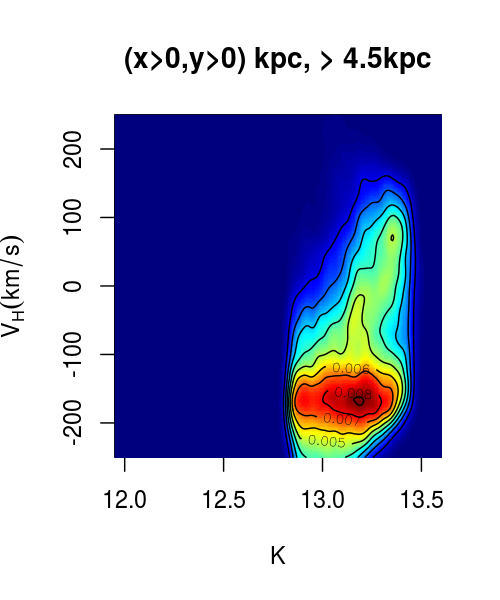}
\hspace{10pt}
\includegraphics[width=4.0cm,angle=0]{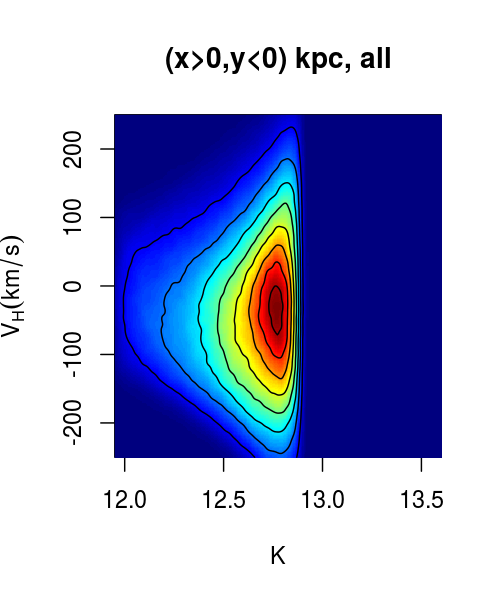}
\includegraphics[width=4.0cm,angle=0]{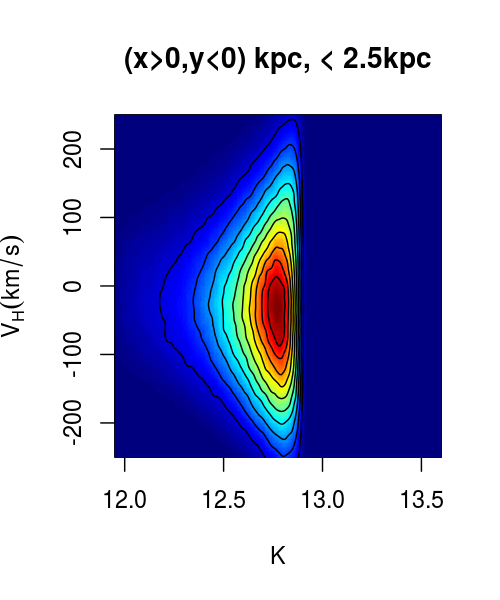}
\includegraphics[width=4.0cm,angle=0]{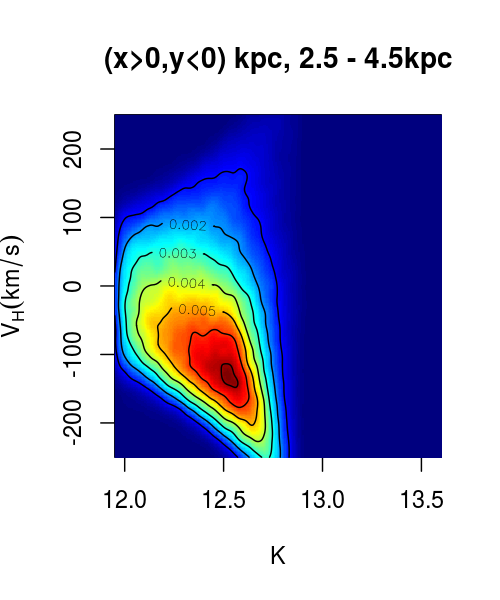}
\includegraphics[width=4.0cm,angle=0]{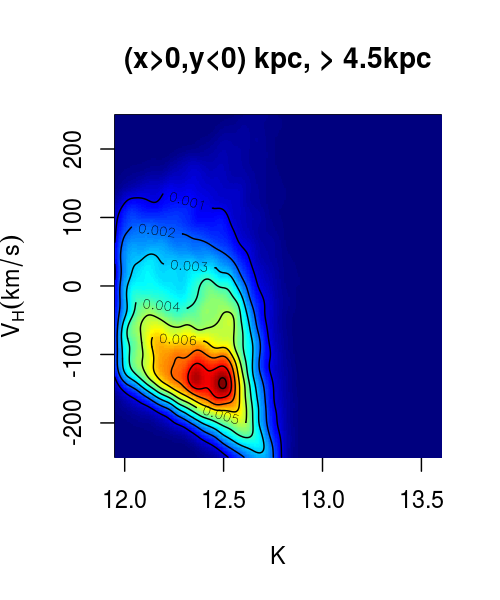}
\hspace{10pt}
\includegraphics[width=4.0cm,angle=0]{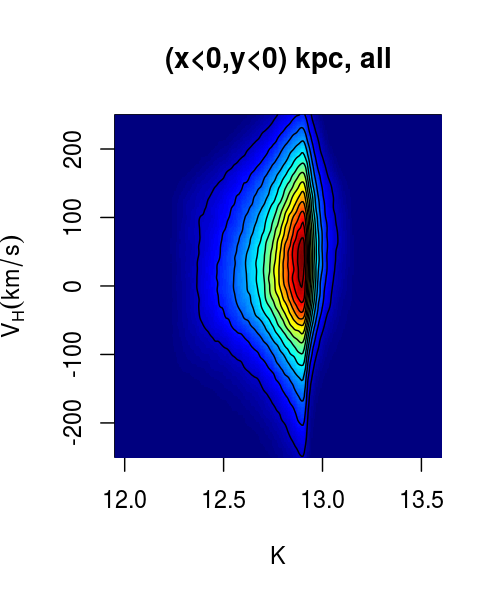}
\includegraphics[width=4.0cm,angle=0]{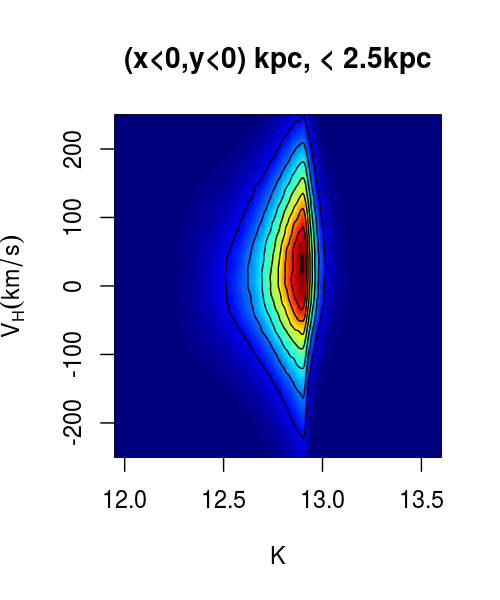}
\includegraphics[width=4.0cm,angle=0]{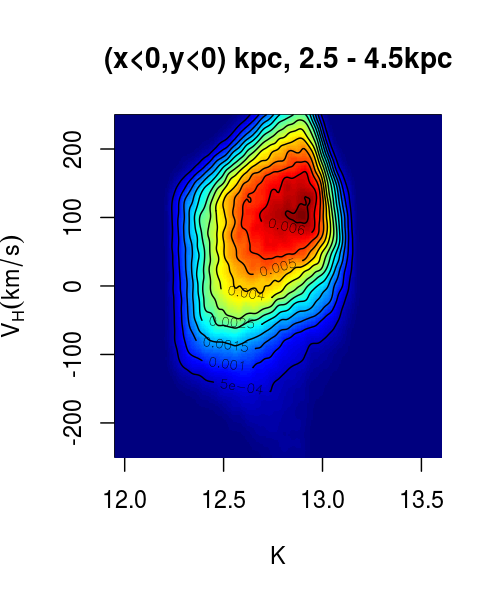}
\includegraphics[width=4.0cm,angle=0]{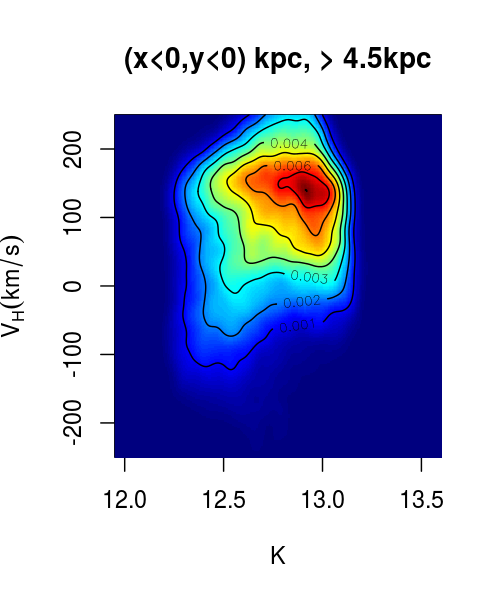}
\caption{$V_H$ versus K-magnitude (as a proxy of the star-Sun distance) density distributions for the whole sample of bulge stars (first row) and for the zones that correspond with the contribution of the $V_{\pi}$ Galactic component field pattern: $x < 0$,~$y>0$~kpc (second row), $x > 0$,~$y>0$~kpc (third row), $x > 0$,~$y<0$~kpc (fourth row), and $x < 0$,~$y<0$~kpc (fifth row) for different birth radii. From left to right: all stars of the modelled galaxy, stars at birth radii $r_{ini}$ $<$ 2.5 kpc, 2.5 kpc ${\le}r_{ini}$ $<$ 4.5 kpc, and $r_{ini}{\ge 4.5}$ kpc. Only disk stars with $|x|\le 2.5$~kpc, $|y|\le 3$~kpc, and $|z|\le 5$~kpc were selected.}\label{Petalesvpi}
\end{figure*}

\begin{figure*}
\centering
\includegraphics[width=4.0cm,angle=0]{FigPetale_PAPER_G_all.png}
\includegraphics[width=4.0cm,angle=0]{FigPetale_PAPER_G_r25.png}
\includegraphics[width=4.0cm,angle=0]{FigPetale_PAPER_G_r2545.png}
\includegraphics[width=4.0cm,angle=0]{FigPetale_PAPER_G_r45.png}
\hspace{10pt}
\includegraphics[width=4.0cm,angle=0]{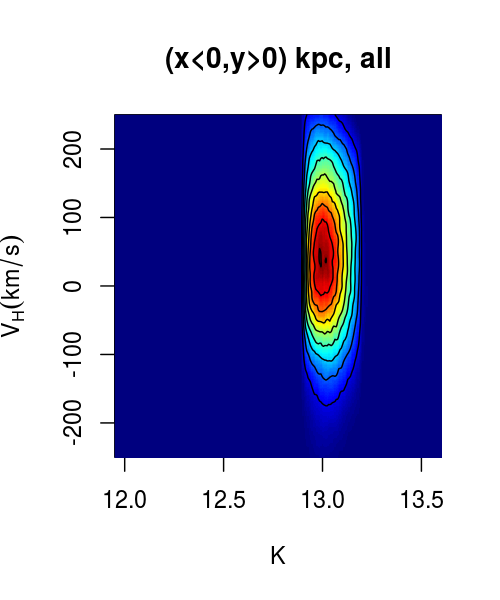}
\includegraphics[width=4.0cm,angle=0]{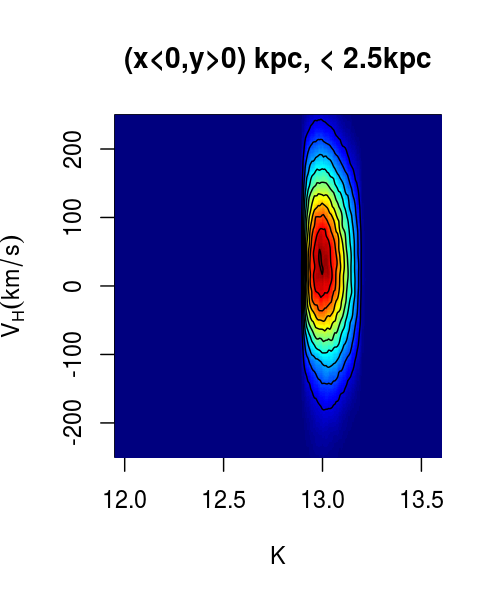}
\includegraphics[width=4.0cm,angle=0]{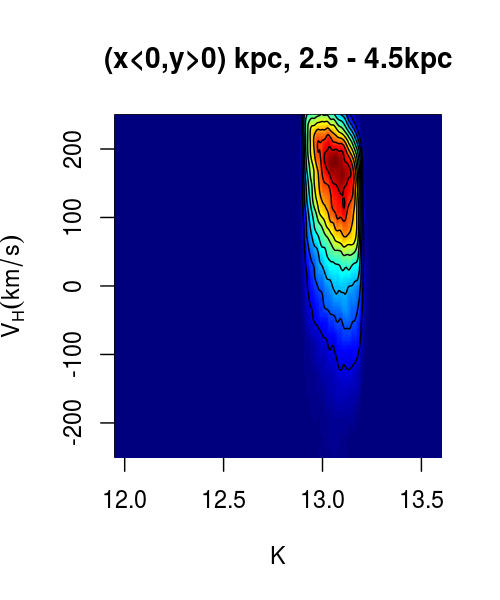}
\includegraphics[width=4.0cm,angle=0]{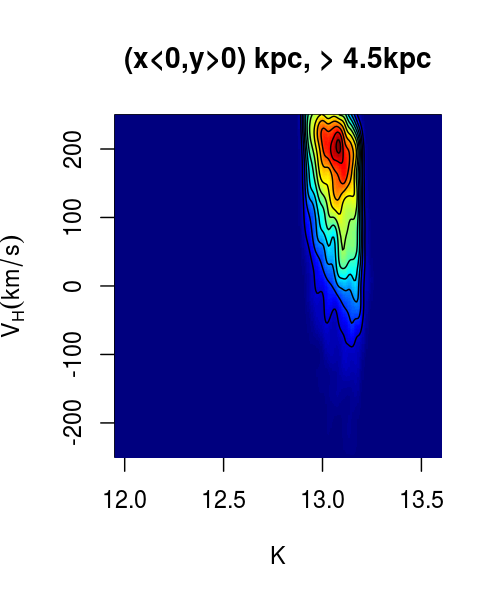}
\hspace{10pt}
\includegraphics[width=4.0cm,angle=0]{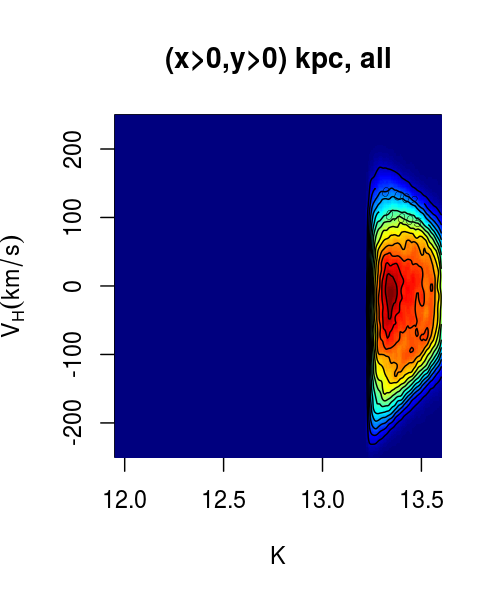}
\includegraphics[width=4.0cm,angle=0]{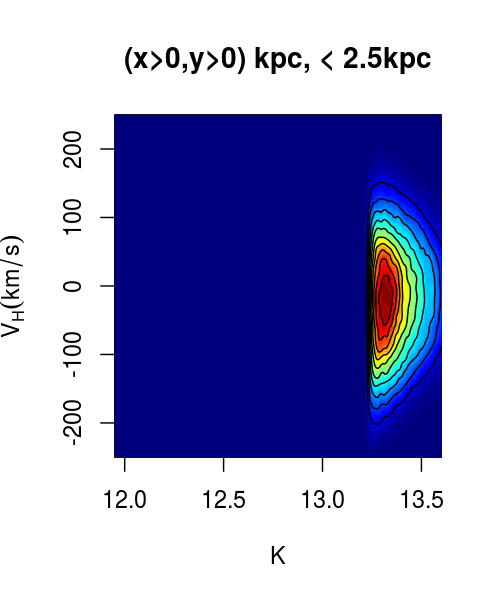}
\includegraphics[width=4.0cm,angle=0]{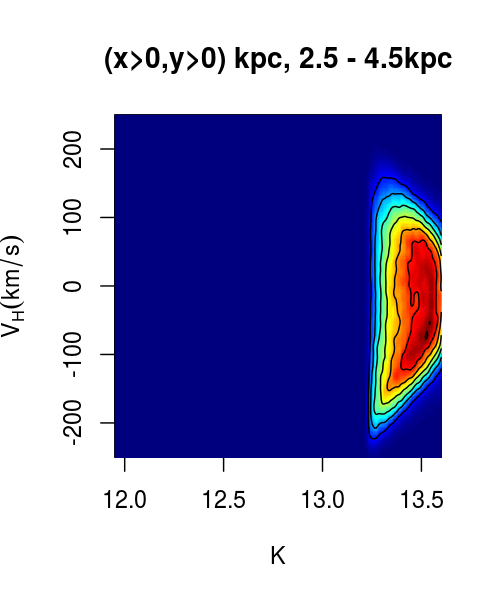}
\includegraphics[width=4.0cm,angle=0]{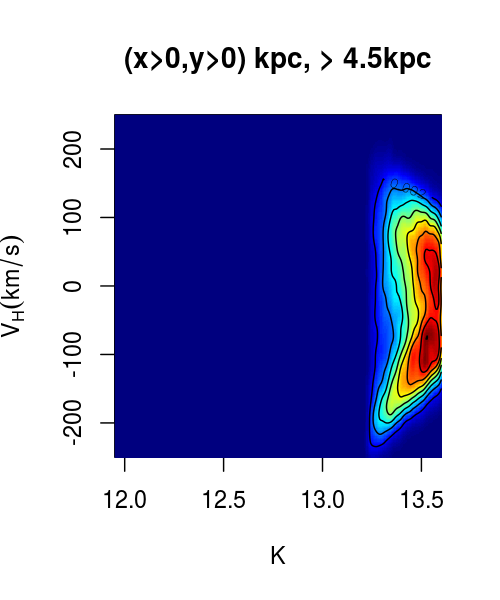}
\hspace{10pt}
\includegraphics[width=4.0cm,angle=0]{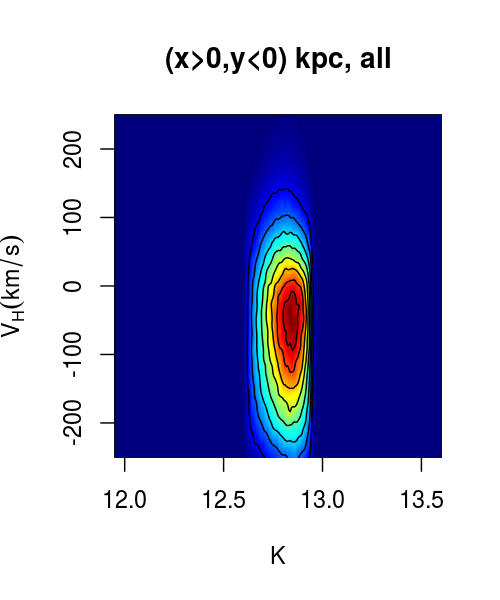}
\includegraphics[width=4.0cm,angle=0]{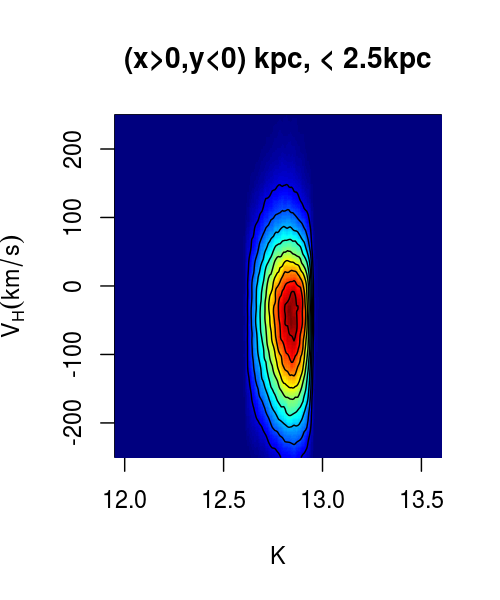}
\includegraphics[width=4.0cm,angle=0]{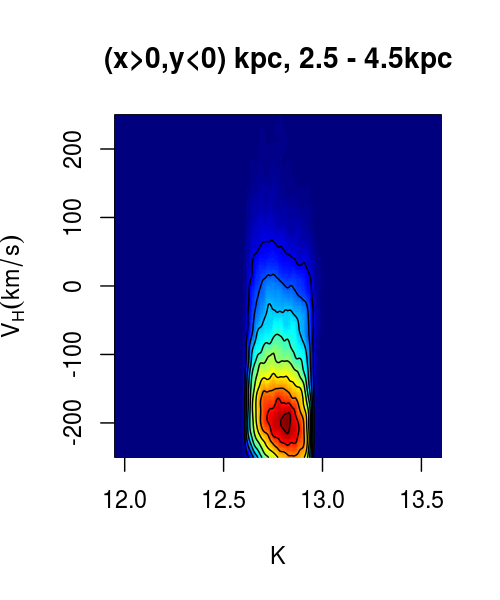}
\includegraphics[width=4.0cm,angle=0]{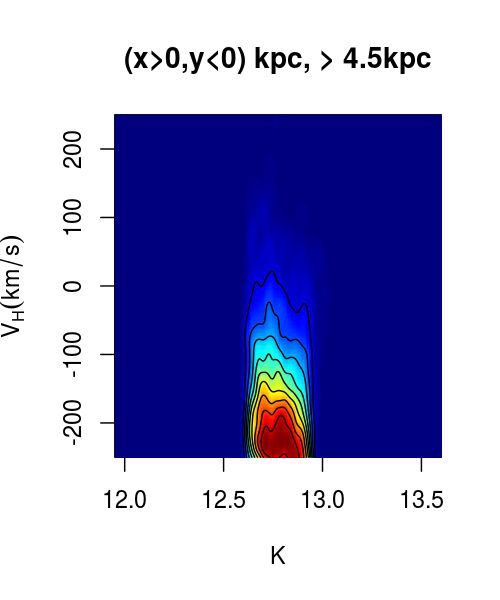}
\hspace{10pt}
\includegraphics[width=4.0cm,angle=0]{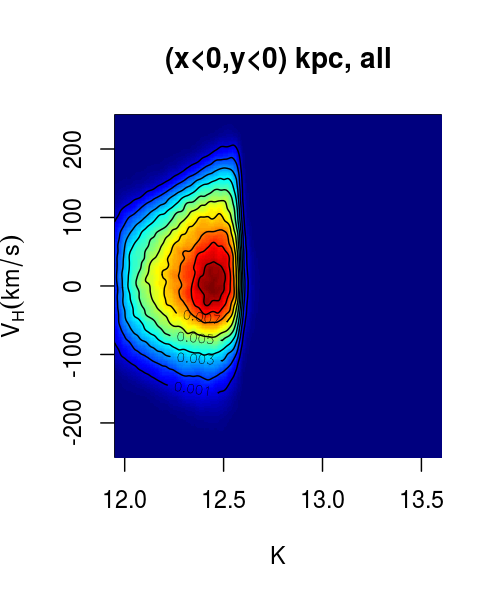}
\includegraphics[width=4.0cm,angle=0]{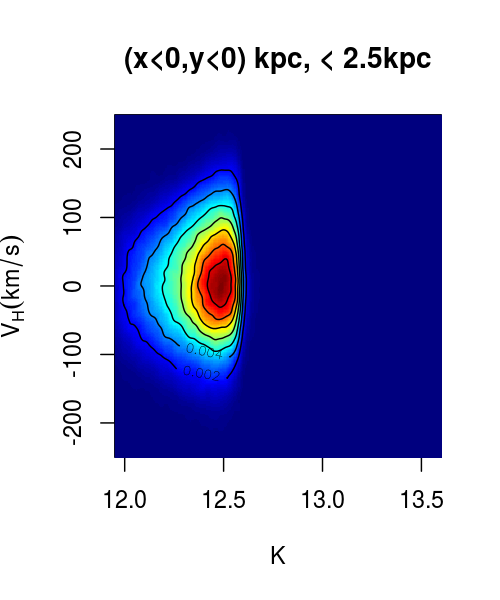}
\includegraphics[width=4.0cm,angle=0]{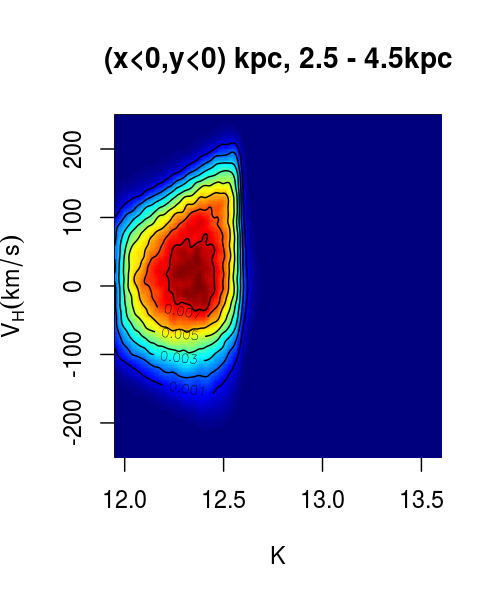}
\includegraphics[width=4.0cm,angle=0]{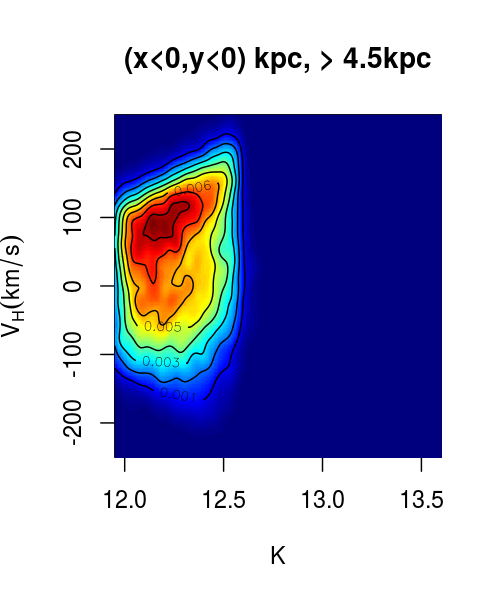}
\caption{$V_H$ versus K-magnitude (as a proxy of the star-Sun distance) density distributions for the whole sample of bulge stars (first row) and for the zones that correspond with the contribution of the $V_{\theta}$ Galactic component field pattern: $x < 0$,~$y>0$~kpc (second row), $x > 0$,~$y>0$~kpc (third row), $x > 0$,~$y<0$~kpc (fourth row), and $x < 0$,~$y<0$~kpc (fifth row) for different birth radii. From left to right: all stars of the modelled galaxy, stars at birth radii $r_{ini}$ $<$ 2.5 kpc, 2.5 kpc ${\le}r_{ini}$ $<$ 4.5 kpc, and $r_{ini}{\ge4.5}$ kpc . Only disk stars with $|x|\le 2.5$~kpc, $|y|\le 3$~kpc, and $|z|\le 5$~kpc were selected. }
\label{Petalesvtheta}
\end{figure*}

The structures in the velocity field shown above should leave a signature on the observed heliocentric radial velocity of the stars in the bulge
region, which should vary with the K-magnitude (or distance). We thus search for possible trends in the $V_H$ versus K-magnitude density distributions that relate to the velocity structures observed in the $V_{\pi}$ and $V_{\theta}$ components velocity fields, as a function of the birth radii. The heliocentric radial velocity, $V_H$, was computed by adopting the circular velocity of the local standard of rest (LSR) at the Sun as 220 $\rm km$ $\rm s^{-1}$ and the solar motion relative to the LSR of 16.5 $\rm km$ $\rm s^{-1}$ toward ($l,b$) = ($53^\circ,25^\circ$)  \citep{mihalas81} as in \cite{beaulieu00}, \cite{kunder12} and \cite{ness13b}. 
In this section, we examine the $V_H$ versus K-magnitude density distributions of the bulge stars integrated over all $|z|$, and leave the analysis of the density distributions in the ($l,b)$ directions that have already been considered in Section~\ref{simall} for the next subsection.\\
First, we isolated the four zones where structures in the velocity field are observed in Figs.~\ref{vpi} and \ref{vtheta}, e.g. those showing outward motions (positive $V_{\pi}$) or inward motions (negative $V_{\pi}$), and near the bar axis, respectively. The delimitation of these regions, different for $V_{\pi}$ and $V_{\theta}$, varies with $r_{ini}$. Different cuts were tested for each velocity component, and the resulting trends on the $V_H$ versus K-magnitude density distributions were similar.
The contribution of each isolated region to the 2D density distributions owing to the observed structures or `patterns` in $V_{\pi}$ and $V_{\theta}$ velocity fields is shown in Figs.~\ref {Petalesvpi} and \ref {Petalesvtheta}, respectively. Only disk stars in the bulge region ($|x|\le 2.5$~kpc, $|y|\le 3$~kpc and $|z|\le5$~kpc) were considered. The figures display the 2D density distributions for the whole sample (first row) and for the four different regions on the Galactic plane that were centred at ($x < 0$,~$y>0$)~kpc (second row), at ($x > 0$,~$y>0$)~kpc (third row), at ($x > 0$,~$y<0$)~kpc (fourth row), and at ($x < 0$,~$y<0$)~kpc (fifth row). Each column corresponds, from left to right, to all stars of the modelled galaxy and to stars at birth radii $r_{ini}$ $<$ 2.5 kpc, 2.5 kpc ${\le}r_{ini}$ $<$ 4.5 kpc, and $r_{ini}$${\ge4.5}$~kpc.  
The 2D density distributions for the whole sample (first row) clearly show that the kinematic characteristics depend on the star birth radii. On the one hand, stars formed in the external galactic regions ($r_{ini}$ $>$ 2.5~kpc), which mainly contribute to the peanut at faint (about K$>$ 13.2) or bright (about K$<$12.6) K-magnitudes (see Section \ref{impX}), may display maxima with high velocities. On the other hand, stars born in the inner galactic regions ($r_{ini}$ $<$ 2.5 kpc), which contribute to the inner part of the peanut, show a global velocity distribution that is quite symmetric with a smaller fraction of high-velocity stars. The $V_H$ velocity dispersions are similar for stars born in the inner and outer disk (115 $\rm km$ $\rm s^{-1}$ and 104 $\rm km$ $\rm s^{-1}$, respectively). We note that, for stars formed at $r_{ini}$ $<$ 2.5 kpc, K-magnitude bimodality is not detected in the 2D density distribution shown here because more than 61$\%$  of the plotted stars are at latitudes $|b|$ $<$ $4^\circ$ where the X-shaped structure is not observed.\\
We now analyse the contribution of the $V_{\pi}$ pattern and the $V_{\theta}$ pattern, as shown in Figs.~\ref {Petalesvpi} and \ref {Petalesvtheta}, respectively, on the global 2D density distributions of the whole sample (see first row). At distances greater than the Galactic centre distance (K$>$12.9 mag.), the contributions to the velocity distributions come from the patterns of zones centred at ($x < 0$,~$y>0$)~kpc (see second row) and, at ($x > 0$,~$y>0$)~kpc (see third row). At distances smaller than the Galactic centre distance (K$<$12.9 mag.), the contributions to the velocity distributions come from zones centred at ($x >0$,~$y<0$)~kpc (see fourth row), and at ($x < 0$,~$y<0$)~kpc (see fifth row).\\
For stars born in the external galactic regions ($r_{ini}$ $>$ 2.5 kpc) and with K$>$12.9 magnitudes, we observe that the maximum of the absolute values of the $V_H$ distributions can be about 100 $\rm km$ $\rm s^{-1}$ or greater. The mean $V_H$ values owing to the $V_{\pi}$ pattern amounts to about 40 $\rm km$ $\rm s^{-1}$ (see second row) and -90 $\rm km$ $\rm s^{-1}$ (see third row) with radial velocity dispersions of about 100 $\rm km$ $\rm s^{-1}$ in both cases. 
The corresponding mean values owing to the $V_{\theta}$ pattern are 118 and 141 $\rm km$ $\rm s^{-1}$ (depending on the birth radius, see second row), and about -28 $\rm km$ $\rm s^{-1}$ (see third row), and the radial velocity dispersions being similar  (about 95 $\rm km$ $\rm s^{-1}$). At distances smaller than the Galactic centre distance, the maximum of the $V_H$ distributions can rise to high values up to 100 $\rm km$ $\rm s^{-1}$ or higher with opposite signs.  
We also observe in the case of the $V_{\theta}$ pattern the presence of more than one maximum. The mean $V_H$ values due to the $V_{\pi}$ pattern are about 75 and 95 $\rm km$ $\rm s^{-1}$ (see fifth row) and -59 and -75 $\rm km$ $\rm s^{-1}$ (see fourth row) for stars that have 2.5 kpc ${\le}r_{ini}$ $<$ 4.5 kpc and $r_{ini}$${\ge4.5}$~kpc, respectively, with a similar radial velocity dispersion of about 95 $\rm km$ $\rm s^{-1}$. The corresponding mean values in Fig.~\ref {Petalesvtheta} are 21 and 32 $\rm km$ $\rm s^{-1}$ and -147 and -182 $\rm km$ $\rm s^{-1}$, and the resulting radial velocity dispersions are of about 90 $\rm km$ $\rm s^{-1}$ in both cases.
Furthermore, several zones contribute to the feature that is observed around K = 12.9 magnitude. The combination of all these contributions produces the trend that is observed in the 2D density distributions of the whole sample for stars that are formed in the external regions.
We emphasise that at K greater than 12.9 magnitudes, the main contribution of the $V_{\pi}$ pattern is observed at K of about 13.2 magnitudes or greater, while in the case of the $V_{\theta}$ pattern, there is a main contribution at positive radial velocities at K smaller than 13.2 magnitudes.\\
For stars formed in the inner regions ($r_{ini}$ $<$ 2.5 kpc) the maximum of the $V_H$ distributions in all the zones is smaller than about 50 $\rm km$ $\rm s^{-1}$ with opposite signs. At distances greater than the Galactic centre distance, the mean $V_H$ values that are due to the $V_{\pi}$ pattern are about 6 $\rm km$ $\rm s^{-1}$ (see second row) and -40 $\rm km$ $\rm s^{-1}$  (see third row), with similar radial velocity dispersions of about 100 $\rm km$ $\rm s^{-1}$. The corresponding values that are due to the $V_{\theta}$ pattern are of about 30 $\rm km$ $\rm s^{-1}$ and -20 $\rm km$ $\rm s^{-1}$ and the radial velocity dispersions amount to 90 $\rm km$ $\rm s^{-1}$ and 75 $\rm km$ $\rm s^{-1}$, respectively. At distances smaller than the Galactic centre distance, the mean $V_H$ values that are due to the $V_{\pi}$ pattern are -25 $\rm km$ $\rm s^{-1}$ (see fourth row) and 22 $\rm km$ $\rm s^{-1}$ (see fifth row), with a dispersion of about 100 $\rm km$ $\rm s^{-1}$ in both cases. The mean  $V_H$ values due to the $V_{\theta}$ pattern are -53 km $\rm s^{-1}$ (see fourth row) and 7 km $\rm s^{-1}$ (see fifth row), with the corresponding radial velocity dispersion of about 90 km $\rm s^{-1}$ and 75 km $\rm s^{-1}$.

In summary, we  analysed the general trends of the signatures that are expected on the observed heliocentric radial velocity $V_H$ versus K-magnitude density distributions, which are due to the velocity structures observed in the $V_{\pi}$, $V_{\theta}$ velocity fields as a function of the star birth radius. The resulting kinematics varies significantly with the star birth radius. Bulge stars that originated in the outer disk show maxima with positive and negative high values that can be larger than 100 $\rm km$ $\rm s^{-1}$. Stars formed in the galactic inner regions display a rather symmetric velocity distribution and a smaller fraction of high-velocity stars. The radial velocity dispersion of the global distributions is slightly higher for stars born in the external regions.\\
The star birth radii imprint on the velocity field as shown in this section should also be present in the 2D density distributions that are observed in different latitude-longitude fields, as discussed in the following.

%
%

\subsection{Heliocentric radial velocity -- K-magnitude density distributions in different directions}
 In this section, we present the 2D density distribution ($V_H$,K ) at latitudes $b=-4^\circ$, $-6^\circ$, $-8^\circ$, and $-10^\circ$ along the bulge minor axis ($l=0^\circ$) as well as outside it, at longitudes $l=\pm5^\circ$ and $l=\pm10^\circ$. Only disk stars with $|x|\le 2.5$~kpc and $|y|\le 3$~kpc were selected. For each observed direction, the line of sight crosses different velocity patterns and, as a consequence, the kinematic signatures on the observed density distributions should change.
 
\begin{figure*}
\centering
\includegraphics[trim = 0cm 0cm 0cm 0cm, width=1\textwidth,angle=0]{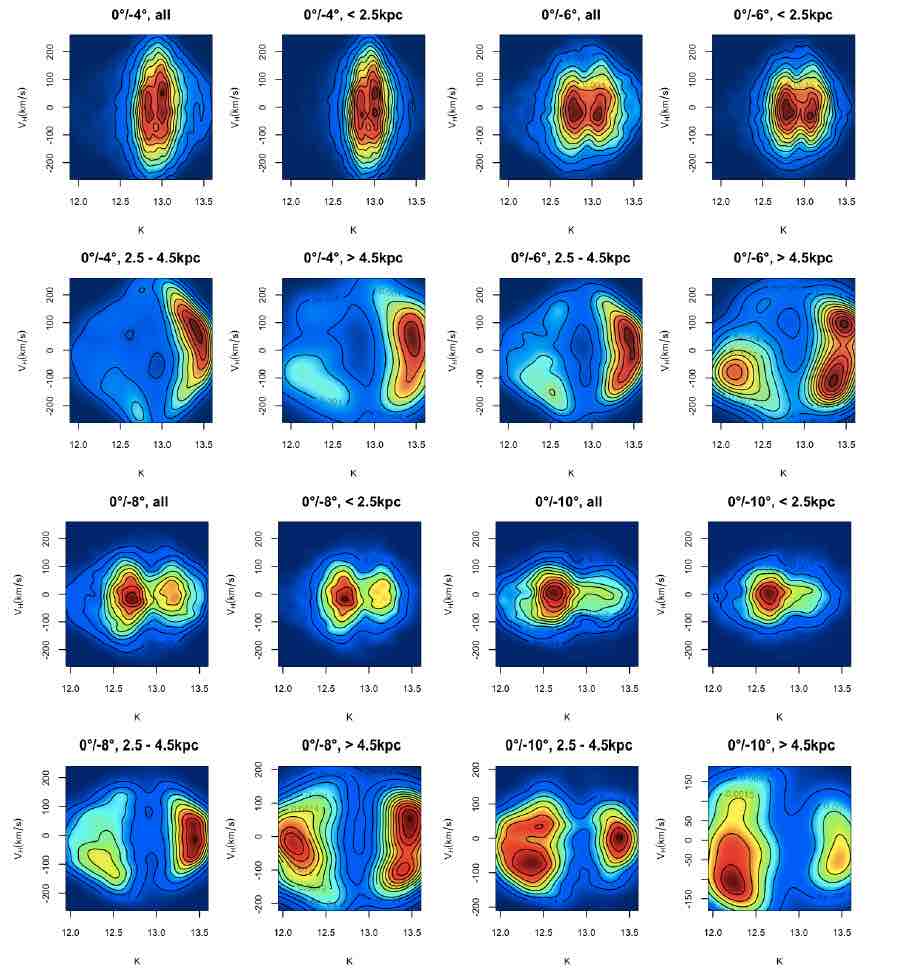}
\caption{Heliocentric radial velocity $V_H$ - K-magnitude (as a proxy of the star-Sun distance) density distributions along the bulge minor axis at $b$=$-4^\circ$, $-6^\circ$, $-8^\circ$, and $-10^\circ$. The size of the fields is $ \Delta{l}=\Delta{b} =1^\circ$ for  $b$ ${\geq}-6^\circ$ and $ \Delta{l}=\Delta{b} =1^\circ.5$ for $b$ $ < -6^\circ$. For each field we show first, all stars of the modelled galaxy, and then stars selected on their birth radii according to: $r_{ini}$ $<$ 2.5 kpc , 2.5  kpc ${\le}r_{ini}$ $<$ 4.5 kpc, and  $r_{ini}$${\ge4.5}$ kpc.
Only disk stars with $|x|\le 2.5$~kpc and $|y|\le 3$~kpc were selected. 
}
\label{VrH_l0}
\end{figure*}

Along the bulge minor axis, the signature on the heliocentric radial velocities comes mainly from the velocity structure that is observed in the $V_{\pi}$ component. Figure~\ref {VrH_l0} shows the heliocentric radial velocity versus K-magnitude density distributions centred at $l=0^\circ$ for the different latitudes. For each latitude, we plotted the density distribution for all stars and for stars selected on their birth radii.\\
Figure~\ref {VrH_l0} clearly shows the K-magnitude bimodality, as found in Section \ref{impX}. For stars formed at $r_{ini}$ $<$ 2.5 kpc, the difference of the mean $V_H$ around each peak is smaller than 15 $\rm km$ $\rm s^{-1}$ at all latitudes, and the radial velocity dispersion decreases with latitude, varying from about 116 $\rm km$ $\rm s^{-1}$ at $b=-4^\circ$ to 68 $\rm km$ $\rm s^{-1}$ at $b=-10^\circ$.
For stars born at $r_{ini}$ $\ge$ 2.5 kpc, we observe the kinematic structures shown in the previous section, with the presence of observed density maxima at positive and/or negative mean $V_H$ velocities and high absolute values, depending on the observed direction.\\
Figure~\ref {VrH_l0} also shows, as expected, that the observed velocity field structures are not symmetric with respect to the Galactic centre. Hence, the difference of the mean $V_H$ velocities varies according to the position of stars, on the near (K$<$12.9 mag.) or on the far (K$>$12.9 mag.) sides of the bar, the so-called bright and faint sides, respectively. Moreover, the variation depends not only on the latitude, but also on the stars birth radii. The obtained mean $V_H$ velocities differences are compiled in Table~\ref{delVrH} which, for each direction and selected sample, presents the number of stars, the mean $V_H$ velocities corresponding to the bright and faint sides, as well as the modulus of its difference ($|\Delta(<V_{H}>)|$). The two main results are: in all selected samples, the obtained difference decreases from $b=-4^\circ$ to $b=-10^\circ$; the difference is smaller than $20$ $\rm km$ $\rm s^{-1}$ for the whole sample and for stars with $r_{ini}$ $<$ 2.5 kpc while, for stars with $r_{ini}$ $>$ 2.5 kpc, the difference can be as high as $40$ $\rm km$ $\rm s^{-1}$ at latitudes greater than $-8^\circ$.

The 2D distribution densities $V_H$ versus K-magnitude outside the bulge minor axis are shown in 
Figs.~\ref{VrH_bm4lno0}, 14, 15, and 16.
Depending on the observed direction, the line of sight crosses only one or two peaks of the X-shape distribution (see Section~\ref{impX}). In general K-bimodality is not found for stars formed at $r_{ini}$ $<$ 2.5 kpc. Moreover, no structure is observed in the velocity field, except for the directions at $l=-5^\circ$ and $b$$<-6^\circ$. On the contrary, stars formed at $r_{ini}$ ${\ge2.5}$ kpc clearly exhibit structures in the 2D density distributions. As expected, observed density maxima at positive and/or negative $V_H$ velocities are obtained. In some cases, the structure of the velocity field is quite complex and several maxima are observed. Table~\ref{delta_vel} gives the modulus of the mean $V_H$ velocity differences between the bright and faint sides for the whole sample in each direction. The values are smaller for positive longitudes (less than about $16$ $\rm km$ $\rm s^{-1}$) than for negative ones (less than $26$ $\rm km$ $\rm s^{-1}$). There is no significant variation of $|\Delta(<V_{H}>)|$ for positive longitudes with the star birth radius while, for negative longitudes, $|\Delta(<V_{H}>)|$ can be as high as $40$ $\rm km$ $\rm s^{-1}$ for stars formed in the external regions.

\begin{figure*}
\centering
\includegraphics[trim = 0cm 0cm 0cm 0cm, width=1\textwidth,angle=0]{Fig12_Quality2.jpeg}
\caption{Heliocentric radial velocity $V_H$ - K-magnitude (as a proxy of the star-Sun distance) density distributions out of the bulge minor axis at $b$=$-4^\circ$ and $l$=$5^\circ$, $-5^\circ$, $10^\circ$, and $-10^\circ$. The size of the fields is $ \Delta{l}=\Delta{b} =1^\circ$ for $b$ ${\geq}-6^\circ$ and $ \Delta{l}=\Delta{b} =1^\circ.5$ for $b$ $ < -6^\circ$. For each field we show first, all stars of the modelled galaxy, and then stars selected on their birth radii according to: $r_{ini}$ $<$ 2.5 kpc, 2.5  kpc ${\le}r_{ini}$ $<$ 4.5 kpc, and  $r_{ini}$${\ge4.5}$ kpc.
Only disk stars with $|x|\le 2.5$~kpc and $|y|\le 3$~kpc were selected. 
}
\label{VrH_bm4lno0}
\end{figure*}

\begin{figure*}
\centering
\includegraphics[trim = 0cm 0cm 0cm 0cm, width=1\textwidth,angle=0]{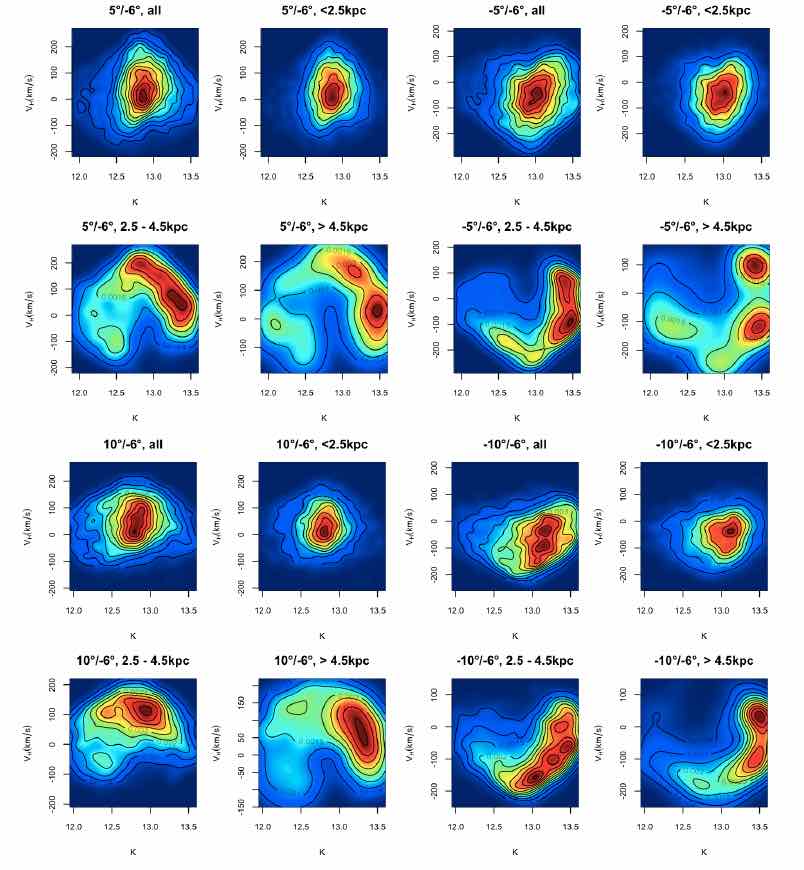}
\label{VrH_bm6lno0}
\caption{Heliocentric radial velocity $V_H$ - K-magnitude (as a proxy of the star-Sun distance) density distributions out of the bulge minor axis at $b$=$-6^\circ$ and $l$=$5^\circ$, $-5^\circ$, $10^\circ$, and $-10^\circ$. The size of the fields is $ \Delta{l}=\Delta{b} =1^\circ$ for  $b$ ${\geq}-6^\circ$ and $ \Delta{l}=\Delta{b} =1^\circ.5$ for $b$ $ < -6^\circ$. For each field we show first, all stars of the modelled galaxy, and then stars selected on their birth radii according to: $r_{ini}$ $<$ 2.5 kpc, 2.5  kpc ${\le}r_{ini}$ $<$ 4.5 kpc, and  $r_{ini}$${\ge4.5}$ kpc. Only disk stars with $|x|\le 2.5$~kpc and $|y|\le 3$~kpc were selected. 
}
\end{figure*}

\begin{figure*}
\centering
\includegraphics[trim = 0cm 0cm 0cm 0cm, width=1\textwidth,angle=0]{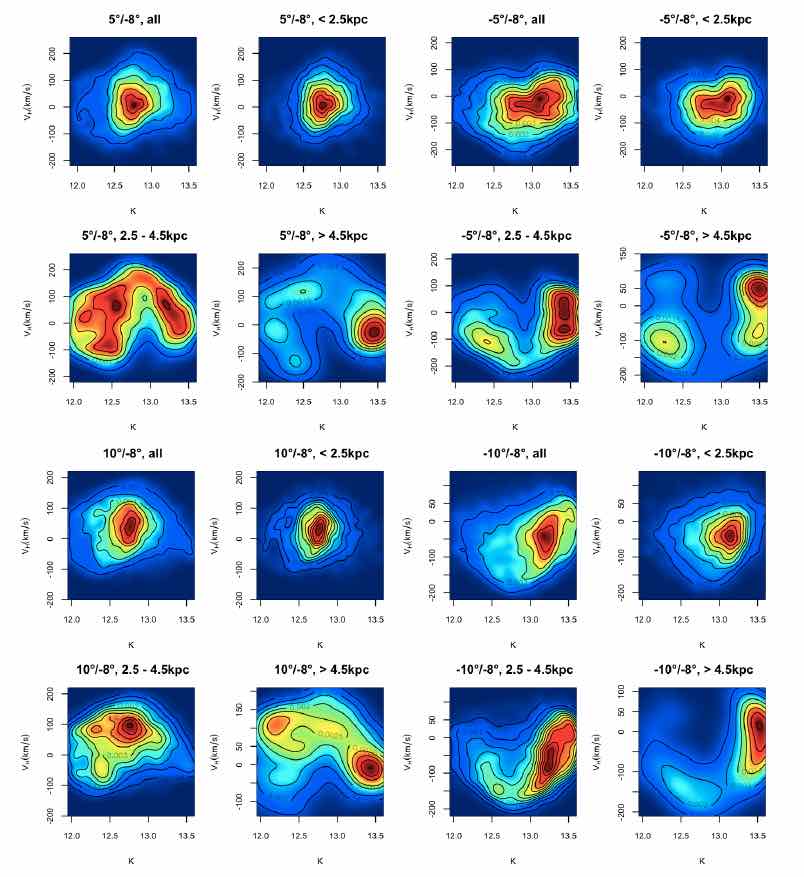}
\label{VrH_bm8lno0}
\caption{Heliocentric radial velocity $V_H$ - K-magnitude (as a proxy of the star-Sun distance) density distributions out of the bulge minor axis at $b$=$-8^\circ$ and $l$=$5^\circ$, $-5^\circ$, $10^\circ$, and $-10^\circ$. The size of the fields is $ \Delta{l}=\Delta{b} =1^\circ$ for  $b$ ${\geq}-6^\circ$ and $ \Delta{l}=\Delta{b} =1^\circ.5$ for $b$ $ < -6^\circ$. For each field we show first, all stars of the modelled galaxy, and then stars selected on their birth radii according to: $r_{ini}$ $<$ 2.5 kpc, 2.5  kpc ${\le}r_{ini}$ $<$ 4.5 kpc, and  $r_{ini}$${\ge4.5}$ kpc.
Only disk stars with $|x|\le 2.5$~kpc and $|y|\le 3$~kpc were selected. 
}
\end{figure*}

\begin{figure*}
\centering
\includegraphics[trim = 0cm 0cm 0cm 0cm, width=1\textwidth,angle=0]{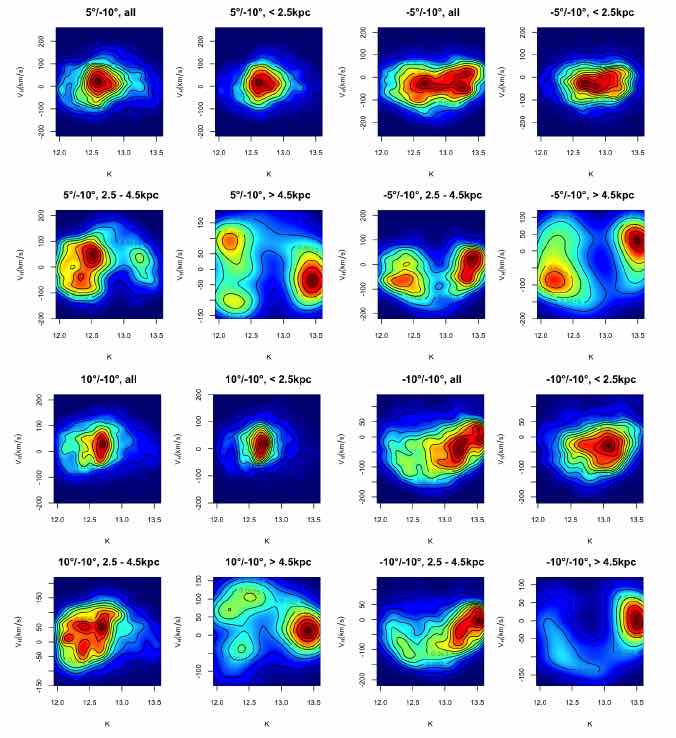}
\label{VrH_bm10lno0}
\caption{Heliocentric radial velocity $V_H$ - K-magnitude (as a proxy of the star-Sun distance) density distributions out of the bulge minor axis at $b$=$-10^\circ$ and $l$=$5^\circ$, $-5^\circ$, $10^\circ$, and $-10^\circ$. The size of the fields is $ \Delta{l}=\Delta{b} =1^\circ$ for  $b$ ${\geq}-6^\circ$ and $ \Delta{l}=\Delta{b} =1^\circ.5$ for $b$ $ < -6^\circ$. For each field we show first, all stars of the modelled galaxy, and then stars selected on their birth radii according to: $r_{ini}$ $<$ 2.5 kpc, 2.5  kpc ${\le}r_{ini}$ $<$ 4.5 kpc, and  $r_{ini}$${\ge4.5}$ kpc.
Only disk stars with $|x|\le 2.5$~kpc and $|y|\le 3$~kpc were selected. 
}
\end{figure*}

\begin{table*}[h]
\caption{Mean heliocentric radial velocity differences between stars at the bright and faint sides of the bar along the bulge minor axis for $b=-4^\circ, -6^\circ, -8^\circ$, and $-10^\circ$ for the whole sample and for stars selected on their birth radii according to: $r_{ini}$ $<$ 2.5 kpc , 2.5  kpc ${\le}r_{ini}$ $<$ 4.5 kpc, and  $r_{ini}{\ge4.5}$ kpc. Only disk stars with $|x|\le 2.5$~kpc and $|y|\le 3$~kpc were selected.}
\label{delta_vel}
\centering
\begin{tabular}{ccrcrcc} 
\hline
Selection & $(l,b)$ & N bright & $<V_{H}> bright$ & N faint & $<V_{H}> faint$ & $|\Delta(<V_{H}>)|$  \\
 & (\degr) &  &  (km/s) &  & (km/s) &  (km/s)  \\
\hline
\hline
 Whole sample & (0,-4)& 25035 & -15.5 $\pm$ 0.8 & 33961 & 1.5 $\pm$ 0.6 & 17.0 $\pm$ 1.0\\
 & (0,-6) & 12003 & -17.1 $\pm$ 0.9 & 12865 & -0.7 $\pm$ 0.8 & 16.4 $\pm$ 1.2\\
 & (0,-8) & 12726 & -15.4 $\pm$ 0.7 & 9215 & -4.1 $\pm$ 0.8 & 11.1 $\pm$ 1.1\\
& (0,-10) & 6094& -12.2 $\pm$ 0.9 & 2877 & -3.7 $\pm$ 1.3 & 8.5 $\pm$  1.6\\
\hline
$r_{ini}$ $<$ 2.5 kpc & (0,-4) & 22445 & -12.5 $\pm$ 0.8 &  28977 & -1.6 $\pm$ 0.7 & 10.9 $\pm$ 1.1 \\
& (0,-6) & 10096 & -12.9 $\pm$ 0.9 & 10205 & -1.4 $\pm$ 0.9 & 11.5 $\pm$ 1.3\\
& (0,-8) & 9711 & -12.2 $\pm$ 0.8 & 6454 & -4.7 $\pm$ 0.9 & 7.5 $\pm$ 1.2\\
& (0,-10) & 4053 & -8.5 $\pm$ 1.1 & 1889 & -4.1 $\pm$ 1.6 & 4.4 $\pm$ 1.9\\
\hline
2.5$<$$r_{ini}$ $<$ 4.5 kpc & (0,-4)& 2423 & -40.8 $\pm$ 3.0 & 4682 & 19.7 $\pm$ 1.8 & 60.5 $\pm$ 3.5 \\
& (0,-6) & 1784 & -38.6 $\pm$ 2.8 & 2512 & 2.8 $\pm$ 2.1 & 41.4 $\pm$ 3.5\\
& (0,-8) & 2842 & -26.0 $\pm$ 1.8 & 2615 & -1.9 $\pm$ 1.7 & 24.1 $\pm$  2.5\\
& (0,-10) & 1915 & -19.3 $\pm$ 1.8 & 933 & -2.5 $\pm$ 2.3 & 16.8 $\pm$  2.9\\
\hline
$r_{ini}$ $>$ 4.5 kpc & (0,-4) & 167 & -46.8 $\pm$ 10.6 & 302 & 10.5 $\pm$7.2 & 57.3 $\pm$ 12.8\\
& (0,-6) & 123 & -50.6 $\pm$ 11.0 & 148 & -10.1 $\pm$ 9.5 & 40.5 $\pm$ 14.5 \\
& (0,-8) & 173 & -22.1 $\pm$ 7.8 & 146 & -15.5 $\pm$ 7.6 & 6.6$\pm$ 11.0\\
& (0,-10) & 126 & -21.9 $\pm$ 8.7 & 55 & -13.4 $\pm$ 10.0 & 8.5 $\pm$ 13.2\\
\label{delVrH}
\end{tabular}
\end{table*}

\begin{table*}[h]
\caption{Mean heliocentric radial velocity differences between stars at the bright and faint sides of the bar in the directions centered at  $l$${\pm5^\circ}$ and ${\pm10^\circ}$ for $b=-4^\circ, -6^\circ, -8^\circ$, and $-10^\circ$ for the whole sample. Only disk stars with $|x|\le 2.5$~kpc and $|y|\le 3$~kpc were selected.}
\label{delta_vel}
\centering
\begin{tabular}{crcrcc} 
\hline
 $(l,b)$ & N bright & $<V_{H}> bright$ & N faint & $<V_{H}> faint$ & $|\Delta(<V_{H}>)|$  \\
 (\degr) &  &  (km/s) &  & (km/s) &  (km/s)  \\
\hline
\hline
(-10,-4)& 4324 & -80.4 $\pm$ 1.2 & 8368 & -61.1 $\pm$ 0.9 & 19.3 $\pm$ 1.5\\
(-10,-6) & 3258 & -75.8 $\pm$ 1.3 & 5999& -52.6 $\pm$ 0.9 & 23.2 $\pm$ 1.6\\
(-10,-8) & 4325 & -65.4 $\pm$ 1.1 & 7578 & -39.4 $\pm$ 0.7 & 26.0 $\pm$ 1.3\\
(-10,-10) & 2221& -51.7 $\pm$ 1.4 & 2819 & -29.0 $\pm$ 1.1 & 22.7 $\pm$  1.8\\
 \hline
(-5,-4)& 12641 & -68.4 $\pm$ 0.9 & 22064 & -53.9 $\pm$ 0.7 & 14.5 $\pm$ 1.1\\
(-5,-6) & 8037 & -64.7 $\pm$ 1.0 & 12499 & -42.9 $\pm$ 0.8 & 21.8 $\pm$ 1.3\\
(-5,-8) & 8828 & -49.6 $\pm$ 0.8 & 10734 & -25.8 $\pm$ 0.7 & 23.8 $\pm$ 1.1\\
(-5,-10) & 4027& -36.8  $\pm$ 1.1 & 3498 & -21.4 $\pm$ 1.1 & 15.4 $\pm$  1.6\\ 
\hline
(5,-4)& 15639 & 32.6 $\pm$ 0.8 & 16750 & 48.0$\pm$ 0.7 & 15.4$\pm$ 1.1\\
(5,-6) & 10627 & 28.4 $\pm$ 0.9 & 8099 & 41.0 $\pm$ 0.9 & 12.6 $\pm$ 1.3\\
(5,-8) & 13205 & 16.4 $\pm$ 0.7 & 6564 & 26.3 $\pm$ 0.9 &  9.9 $\pm$ 1.1\\
(5,-10) & 6543 & 8.7  $\pm$ 0.9 & 2265 & 14.2 $\pm$ 1.4 & 5.5 $\pm$ 1.7\\ 
\hline
(10,-4)& 6620 & 47.9 $\pm$ 1.1 & 4768 & 56.5 $\pm$ 1.1 & 8.6 $\pm$ 1.6\\
(10,-6) & 5354 & 39.4 $\pm$ 1.1 & 2815& 46.0 $\pm$ 0.9 & 6.6 $\pm$ 1.4\\
(10,-8) & 8892 & 30.6 $\pm$ 0.7 & 3121 & 31.3 $\pm$ 1.2 & 0.7 $\pm$ 1.4\\(10,-10) & 5632 & 19.4 $\pm$ 0.8 & 1488 & 22.6 $\pm$ 1.6 & 3.2   $\pm$ 1.8\\
\end{tabular}
\end{table*}

\section{Discussion}
The results obtained in the previous section show that the initial (i.e. before bar formation) location in the disk leaves an imprint in the kinematics of stars. In particular, the presence of stars coming from the external disk may have an impact on the interpretation of some recent observational findings, as explained below.

\subsection{Streaming motions induced by the bar}
We first compare the obtained bright-faint sides differences in mean heliocentric radial velocities, ($|\Delta(<V_{H}>)|$, to observations. As was noted in the introduction, the observed values of these differences are not always consistent with each other.  Along the bulge minor axis, stars show a streaming motion $|\Delta(<V_{H}>)|$ smaller than about $20$$\pm2$ $\rm km$ $\rm s^{-1}$ which decreases with latitude (see Table~\ref{delVrH}). These results are, within the errors, in good agreement with most of the values found in the literature that were obtained from radial velocity measurements: \cite{ness12} ($30$$\pm12$ $\rm km$ $\rm s^{-1}$) at $l=0^\circ$ and $b=-5^\circ$, \cite{vasquez13} ($\sim23$$\pm10$ $\rm km$ $\rm s^{-1}$) at $l=0^\circ$ and $b=-6^\circ$, \cite{depropris11} ($12$$\pm10$ $\rm km$ $\rm s^{-1}$) at  $l=0^\circ$ and $b=-8^\circ$, \cite{uttenthaler12} ($5.2$$\pm9.5$ $\rm km$ $\rm s^{-1}$) at $l=0^\circ$ and $b=-10^\circ$,  \cite{ness12} ($7$$\pm9$ $\rm km$ $\rm s^{-1}$) at $l=0^\circ$ for the combined fields at $b -7.5^\circ$ and $-10^\circ$). In Baade's window \cite{babusiaux10}, selecting the stars on the near and far sides of the bar as those in the first and last  25$\%$ quantiles of the I-magnitude distribution, obtained a higher value of $70$$\pm30$ $\rm km$ $\rm s^{-1}$. 
This selection favours bright and faint stars at large distances from the Galactic centre which, according to our model, should preferentially be stars that were originally located in the outer disk and migrated in the inner region at the time of the bar formation (see \cite{dimatteo14}, Fig. 3). And indeed, for stars with $r_{ini}$${\ge2.5}$~kpc, our model predicts a difference in the mean heliocentric radial velocities of about  $60$$\pm5$ $\rm km$ $\rm s^{-1}$ (see Table~\ref{delVrH}), which is consistent with the value found by \cite{babusiaux10}.\\
Outside the bulge minor axis, at $b=-3.5^\circ$, the results of \cite{rangwala09}  ($40$$\pm11$ $\rm km$ $\rm s^{-1}$ at $l=5.5^\circ$ and $32$$\pm11$ $\rm km$ $\rm s^{-1}$ at  $l=-5.5^\circ$) are, within the errors, compatible with our results.
\subsection{High-velocity stars in the bulge}
\begin{figure}
\centering
 \includegraphics[trim = 0cm 0cm 0cm 0cm, scale= 0.35]{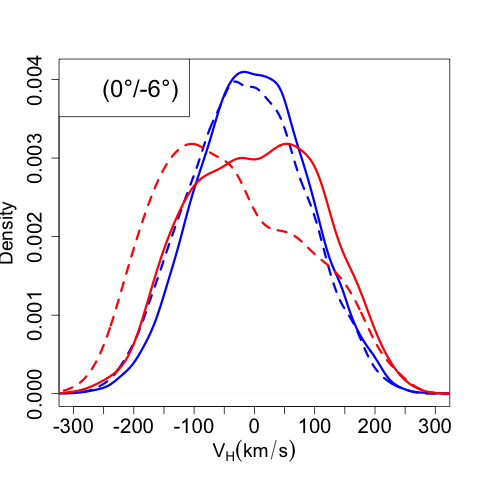}
\caption{Heliocentric radial velocity density distributions in the field centred at $l$=$0^\circ$ and $b$=$-6^\circ$ for stars in the near and far sides of the bar. Only disk stars with $|x|\le 2.5$~kpc and $|y|\le 3$~kpc were selected. The size of the field is $ \Delta{l}=\Delta{b} =1^\circ$. Curves in blue and in red correspond to stars born at $r_{ini}$ $<$ 2.5 kpc and at 2.5  kpc ${\le}r_{ini}$ $<$ 4.5 kpc, respectively, at the bright (dotted line) and faint (solid line) sides of the bar.}\label{vasquez}
\end{figure}

In their study, based on bulge giants in the bright and the faint RC of a field at $l=0^\circ$ and $b=-6^\circ$, \cite{vasquez13} report an excess of stars at $V_{H}$ $\sim$$\pm80$ $\rm km$ $\rm s^{-1}$ in the faint and the bright samples, respectively. This observational result has been interpreted as the presence, in both the bright and faint sides, of stars on elongated orbits: indeed an excess of stars approaching the Sun  should be expected at the near side and an excess of stars receding from the Sun should be observed at the far side. As shown in Fig.~\ref {VrH_l0}, at $(l,b)=(0^\circ,-6^\circ)$, the velocity field of stars that originated in the outer disk displays structures at high positive and negative $V_{H}$, which are not observed among stars formed in the inner disk. In Fig.~\ref {vasquez}, the heliocentric radial velocity distribution is plotted for stars born in the inner and external regions at the bright and faint sides of the bulge.  As we can observe in the case of stars formed in the inner regions (blue curve), the $V_{H}$ density distribution at both sides of the bar is quite symmetric (the mean and the median radial velocity values are close) and high-velocity stars are present in the tails of the distribution. Stars born in the external regions show overdensities (red curve) with high-velocity values (greater than about 50 km $\rm s^{-1}$), which are positive for stars in the faint side and negative for those in the bright side, as observed by \cite{vasquez13}. We conclude that the imprint of stars that are born in the external regions may explain the observed results of \cite{vasquez13}.\\

\begin{figure*}
\centering
 \includegraphics[width= 5.5cm,angle=0]{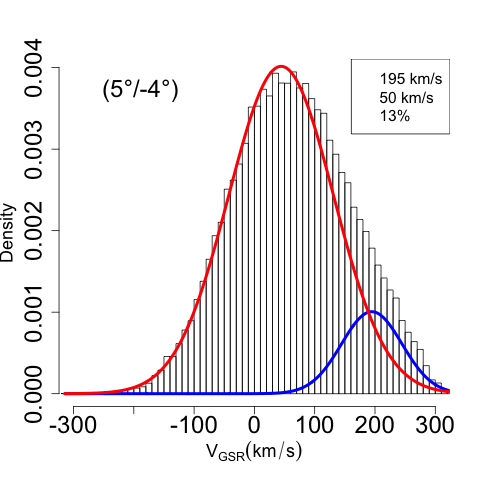}
 \includegraphics[ width= 5.5cm,angle=0]{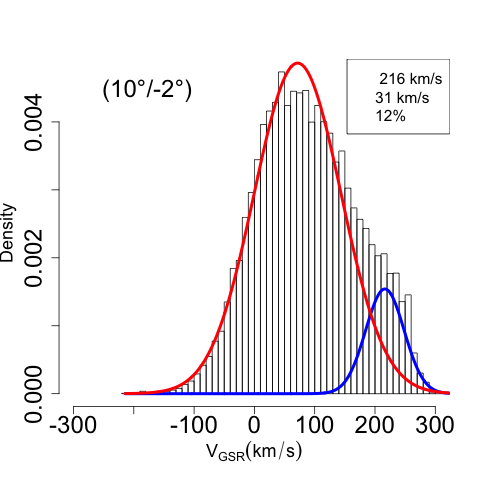}
 \includegraphics[width= 5.5cm,angle=0]{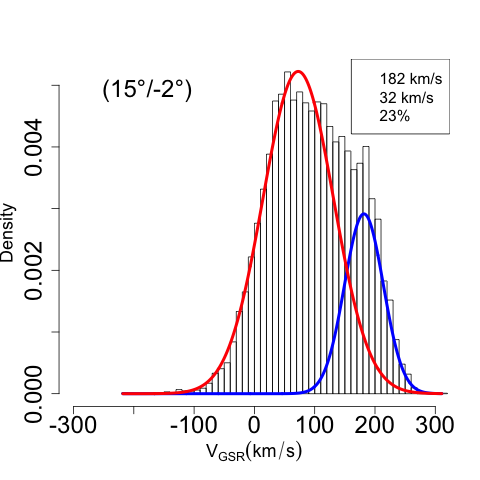}
 \hspace{10pt}
 \includegraphics[width= 5.5cm,angle=0]{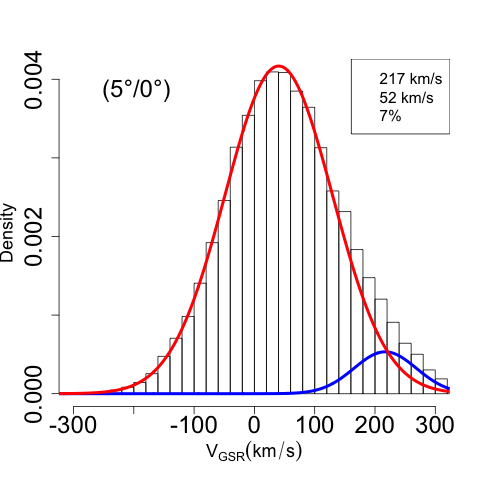}
 \includegraphics[width= 5.5cm,angle=0]{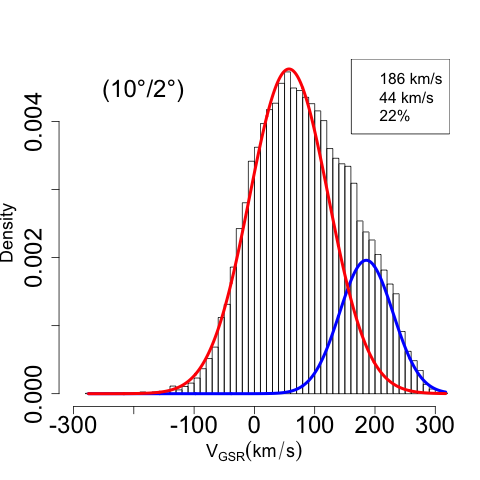}
 \includegraphics[width= 5.5cm,angle=0]{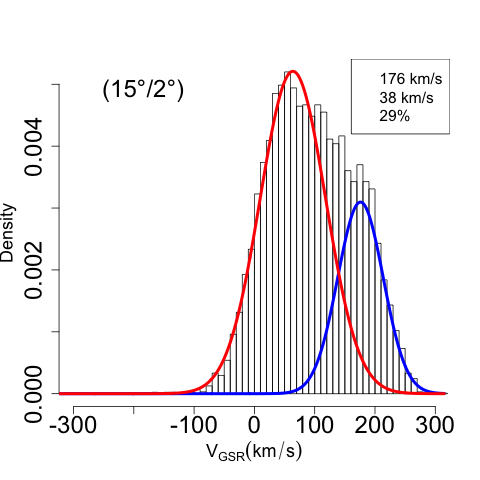}
\caption{ $V_{GSR}$ radial velocity histograms in different fields $(l,b)$ (top left).  The size of each field is $ \Delta{l}=\Delta{b} =1^\circ$ for $l $ $< 15^\circ$ and $1.5^\circ$ otherwise. All stars along each direction were selected. A two-Gaussian model is fitted to the data. The cold kinematic component is in blue and the corresponding mean value, dispersion, and fraction of stars (in percentage) are given at the top right. The mean errors are smaller than 1 km $\rm s^{-1}$ for the mean and the dispersion values and smaller than 1$\%$ for the fraction of stars.}
\label{nidever}
\end{figure*}

\cite{nid12} observe a cold high radial velocity peak ($V_{GSR}$ ${\sim200}$ km $\rm s^{-1}$,  $\sigma_V$ ${\sim30}$ km $\rm s^{-1}$) in the APOGEE commissioning data of Galactic bulge stars, which corresponds to ${\sim10\%}$ of stars in many of the observed fields. The existence of this peak has been questioned by \cite{li14} who do not find a statistically significant cold high-velocity peak in their N-body models. Recently, \cite{moll15} suggest that resonant orbits that show different kinematic features may be used to explain the high-velocity peak. Moreover, the simulation analysed by \cite{aumer15} reproduces the high-velocity feature at latitudes $|b|$$<2^\circ$ and suggests that it is made up preferentially of young bar stars. 
We use our simulation to investigate this issue. We  considered similar fields to those 
where a dual-peak structure has been seen, centred at (l,b): $(5^\circ,-4^\circ)$, $(5^\circ,0^\circ)$, $(10^\circ,\pm2^\circ),$ and $(15^\circ,\pm2^\circ)$. The size of each field is $ \Delta{l}=\Delta{b} =1^\circ$ for $l< 15^\circ$ and $1.5^\circ$ otherwise. 
Following \cite{nid12}, we analyse the radial velocity distribution of $V_{GSR}$ instead of $V_{H}$, fitting a two-Gaussian model to our simulated data. All the stars along the line of sight were selected. The results are shown in Fig.~\ref {nidever}. For the obtained cold component (in blue), the values of the mean, of the dispersion, and of the fraction of stars are given at the top right. In all the regions, we detect the presence of a high-velocity component. The means, the dispersions 
and the fraction values vary between about 180 and 220 $\rm km$ $\rm s^{-1}$, 30 and 50 $\rm km$ $\rm s^{-1}$ and 10 and 30 $\%$, respectively. They are in good agreement with the results of \cite{nid12}. 
Our results confirm the possible existence of high-velocity peaks in the radial velocity distributions. However, the velocity distributions do not show the trough at intermediate velocities ($V_{GSR}$ ${\sim 140 -180}$ km $\rm s^{-1}$), as seen in the observational data whose origin is difficult to interpret \citep{nid12}. 

\subsection{Complex observations versus simple models}
The global 2D density distributions (heliocentric radial velocity versus distance) show a mix of stars with different kinematic characteristics. As shown, at large distances from the Galactic centre distance (about K  $>$13.2 mag and K$<$12.6 mag for RC stars), stars formed in the external regions of the disk largely contribute to the observed distributions. Stars born in the inner regions, which are numerous, are also present but with different kinematic characteristics. We observe the existence of structures and/or high-velocity peaks in the distributions, depending on the observed direction. 
With real data the picture should be more complex. We have only considered the bulge formed from the disk via the resonance with a bar. But classical bulge stars and/or thick disk stars should be present in the bulge. Only few classical bulge stars are expected \citep{shen10, kunder12, dimatteo14} and they do not participate in the B/P-shape bulge. However, our simulation does not include a thick disk component, whose stars may also contribute to the B/P shape (Di Matteo et al, in preparation). The existence of a massive thick disk component, as suggested in the literature \citep{fuhrmann12, haywood13, snaith14, haywood14a, haywood14b, haywood15} should strengthen our results since we expect similar structures in the velocity field, but with a higher dispersion. To discriminate among the different stellar populations, the combination of kinematics, distances, and $\alpha$-elements data is required. At present, it is not possible to make a comparison of the 2D density distribution, as shown in this paper, with real data because it would require large samples of bulge stars with accurate distances. Finally, we note that, according to our results, the presence of substructures, peaks, and clumps in the bulge velocity fields is not necessarily a sign of past accretion events.

\section{Conclusions}
We analysed the imprints of the star birth radii on the bulge kinematics using an N-body simulation of the bulge formed via secular evolution, consisting of an isolated stellar disk, with a B/D=0.1 classical bulge, and containing no gas. As classical bulge stars are not affected by the bar instability mechanism, our results apply to the disk component. Four different latitudes were considered: $b=-4^\circ$, $-6^\circ$, $-8^\circ$, and $-10^\circ$, along the bulge minor axis as well as outside it for $l=\pm5^\circ$ and $l=\pm10^\circ$. Because the velocity field changes with galactic position, we investigated the imprints of the star birth radii on the 2D density distribution given by the heliocentric radial velocity versus K-magnitude, which is a proxy of the heliocentric distance for RC stars.
\begin{enumerate}[(i)]
\item
The imprints of the stars' birth radii are clearly found in the X-shaped structure [Section \ref{impX}]. Stars in the X-shaped structure, which were formed further out in the disk, show larger separation between their peaks, compared to stars that originated in the inner disk. 
As a consequence, the main contribution to the 2D density distributions of stars formed in the external regions appears at the outer bar region.
Stars born in the inner disk, the most numerous, are present at all distances in the bulge and contribute predominantly to the X-shaped structure.
\item
The kinematics of stars varies significantly with their birth radius [Section 4]. The barred potential induces non-circular motions and, as a consequence, the Galactocentric radial and tangential velocity component fields are complex and exhibit some structures. We have analysed the general trend of the signatures left by these structures on the 2D density distributions. 
Stars that originated in the external disk predominantly contribute to the velocity distributions at large distances from the Galactic centre. The resulting velocity distributions of these stars 
may show peaks at positive and negative heliocentric radial velocities with large absolute magnitudes ($>$ 100 $\rm km$ $\rm s^{-1}$), depending on the observed direction. In some cases, particularly outside the bulge minor axis, several peaks are observed. Stars born in the inner disk have different kinematic characteristics to those of the outer disk. They display a rather symmetric velocity distribution and a smaller fraction of high-velocity stars. 
\item
Stellar stream motions are induced by the bar [Sections 4.2 and 5.1]. Along the bulge minor axis, stars formed in the inner disk show a streaming motion that is smaller than about $20$ $\rm km$ $\rm s^{-1}$. This streaming motion decreases as latitude decreases (from $b=-4^\circ$ to $b=-10^\circ$). Stars formed in the external disk show a streaming motion larger than $40$ $\rm km$ $\rm s^{-1}$ at latitudes greater than $b=-8^\circ$. Outside the bulge minor axis, for positive longitudes, there is no significant variation of the streaming motion with the star birth radius ($< $16 $\rm km$ $\rm s^{-1}$). For negative longitudes, the streaming motion can be larger than $40$ $\rm km$ $\rm s^{-1}$ for stars coming from the external disk.
\item
Our results show the existence of high-velocity stars in the bulge that are part of the X-shaped structure. The presence of stars coming from the external disk may explain the cold high radial velocity peak observed in the APOGEE commissioning data \citep{nid12}, as well as the excess of high-velocity stars in the near and far arms of the X-shaped structure at $l$=$0^\circ$ and $b$=$-6^\circ$, as quoted in \cite{vasquez13} [Section 5.2]. 
\item
Our results also indicate that the bulge contains various stellar populations with different kinematic characteristics [Section 5.3]. This work considers only the bulge formed via secular evolution. The kinematics of the bulge stars depends on the initial origin of the stars in the disk and, significantly, varies with the star birth radius.
With real data, the kinematic picture becomes more complex owing to the possible presence of classical bulge stars and/or thick stars in the samples. Moreover, the presence of substructures, peaks, and clumps in the observed bulge velocity fields is not necessarily a sign of past accretion events. To discriminate among the different stellar populations, large samples of bulge stars with corresponding kinematics, distances, and abundance measurements are required. These measurements are expected from upcoming surveys, including APOGEE-2 and Gaia. 
\end{enumerate}

\section*{Acknowledgments} 
We thank the anonymous referee for a very constructive report that greatly helped to improve the paper.

\end{document}